\documentclass{elsart}
\usepackage{amssymb}          
\usepackage[dvips]{graphicx}  

\begin{document} 

\begin{frontmatter} 
\title{Wavelength Scaling and Square/Stripe and Grain Mobility
Transitions in Vertically Oscillated Granular Layers}
\author[nwu]{Paul B. Umbanhowar\thanksref{pbu}} and
\author[texas]{Harry L. Swinney\thanksref{hls}}
\address[nwu]{Department of Physics and Astronomy, Northwestern University\\ 
Evanston, IL 60208}
\address[texas]{Center for Nonlinear Dynamics and Department of
Physics\\ The University of Texas at Austin\\ Austin, TX 78712}
\thanks[pbu]{email: umbanhowar@nwu.edu}
\thanks[hls]{email: swinney@chaos.ph.utexas.edu}

\vspace{-0.2in}
\begin{abstract}
Laboratory experiments are conducted to examine granular wave patterns
near onset as a function of the container oscillation frequency $f$
and amplitude $A$, layer depth $H$, and grain diameter $D$.  The
primary transition from a flat grain layer to standing waves occurs
when the layer remains dilated after making contact with the
container.  With a flat layer and increasing dimensionless peak
container acceleration $\Gamma = 4 \pi^2 f^2 A/g$ ($g$ is the
acceleration due to gravity), the wave transition occurs for $\Gamma
\approx 2.6$, but with decreasing $\Gamma$ the waves persist to
$\Gamma=2.2$.  For $2.2 < \Gamma < 3.8$, patterns are squares for $f <
f_{ss}$ and stripes for $f > f_{ss}$; $H$ determines the square/stripe
transition frequency $f_{ss} = 0.33 \sqrt{g/H}$. The dispersion
relations for layers with varying $H$ collapse onto the curve
$\lambda/H = 1.0 + 1.1 (f \sqrt{H/g})^{-1.32 \pm 0.03}$ when the peak
container velocity $\mathrm{v} = 2 \pi A f$ exceeds a critical value,
$\mathrm{v}_{gm} \approx 3 \sqrt{Dg}$.  Local collision pressure
measurements suggest that $\mathrm{v}_{gm}$ is associated with a
transition in the horizontal {\it g}rain {\it m}obility: for
$\mathrm{v} > \mathrm{v}_{gm}$, there is a hydrodynamic-like
horizontal sloshing motion, while for $\mathrm{v} < \mathrm{v}_{gm}$,
the grains are essentially immobile and the stripe pattern apparently
arises from a bending of the granular layer.  For $f$ at
$\mathrm{v}_{gm}$ less than $f_{ss}$ and $\mathrm{v} <
\mathrm{v}_{gm}$, patterns are tenuous and disordered.
\end{abstract} 
\end{frontmatter} 

\vspace{-0.35in}

\section{Introduction}
\label{intro}

\begin{figure}[t]  	
\centerline{\includegraphics[width=\textwidth]{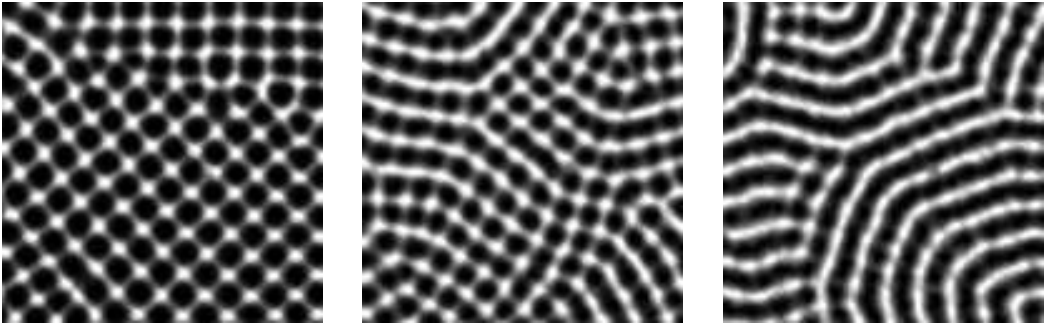}}
  \caption{Images of granular wave patterns near onset: (left)
  squares, $f=33$ Hz ($f^*=0.28$), (center) squares and stripes,
  $f=37$ Hz ($f^*=0.31$), and (right) stripes, $f=43$ Hz
  ($f^*=0.36$). In each case $D=0.17$ mm bronze, $\Gamma=2.5$, $N=4$,
  container is {\it S1}, and the image size is $40.5 \times 40.5$ mm.}
  \label{fig1}
\end{figure}

Granular materials are collections of discrete solids for which even
the simplest realizations --- ensembles of identical solid spheres
interacting only via contact forces --- exhibit a wealth of surprising
behaviors \cite{jnb96}.  Our work on granular media has focused on a
dynamic phenomenon, the formation of subharmonic standing waves in
vertically oscillated granular layers \cite{melo94,melo95}.  These
strongly nonlinear waves form patterns of stripes, squares, hexagons
and more complex patterns \cite{umb97}, as well as localized
structures called oscillons \cite{umb96}, as a function of three
dimensionless control parameters: the acceleration $\Gamma$, the
frequency $f^* = f \sqrt{H/g}$, and the layer thickness $N=H/D$.
This paper concerns the square or stripe patterns that arise at the
primary instability as $\Gamma$ is increased; examples of these
patterns are shown in Fig.~\ref{fig1}.  Despite being composed of
discrete grains and having typical wavelengths $\lambda$ of only 20-30
$D$, the appearance of granular patterns is similar to that of
patterns in fluid systems \cite{cross93,bruyn}, the closest example
being standing surface waves in a vertically oscillated liquid layer
(the Faraday instability) \cite{gollub}.

Pattern formation in granular media can perhaps be described by
continuum equations analogous to those used in fluid system
\cite{jenkins,bizjsp,bizpre}, but the continuum equations remain
largely untested by experiment.  Aspects of granular patterns have
been described using phenomenological models \cite{dyn,amp,map}, but
making the connection to real systems requires an understanding of
granular media at a more microscopic level.  Recently, molecular
dynamics simulations have begun to yield new details concerning the
behavior of granular waves \cite{other_sim}.  In particular, a
simulation developed by Bizon {\it et al.} \cite{bizon98}
quantitatively reproduces the granular patterns observed in three
dimensions and thus allows calculation of quantities not readily
accessible by experiment.

In this paper, we will first describe our experimental system and then
discuss some generic features of the flat-layer/wave transition.
Next, we will present results for the scaling of the square/stripe
transition frequency and the wavelength with particle size and layer
depth.  Last, we will use results from local measurements of the
collision pressure and from dispersion relations to demonstrate and
characterize the grain mobility transition.

\section{Experiment}

\begin{figure}[t]
\centerline{\includegraphics[width=0.5\textwidth]{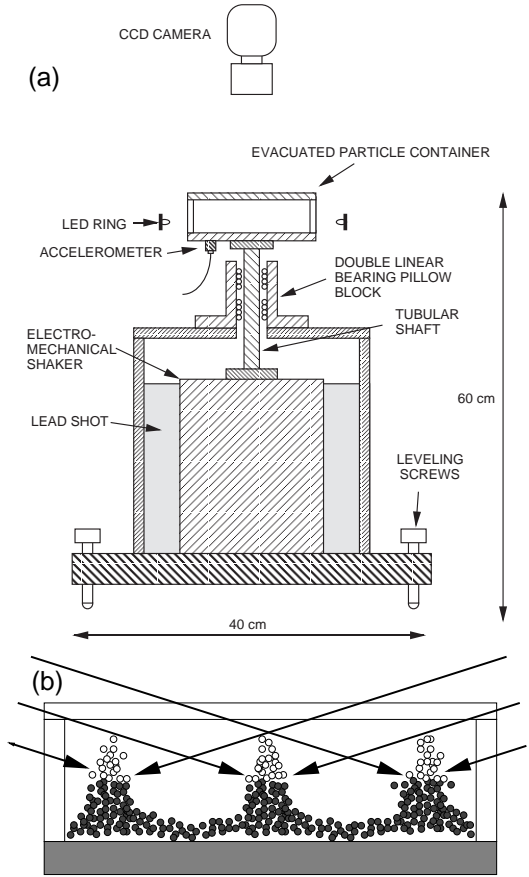}}
\caption{(a) Experimental apparatus and (b) lighting.  Patterns are
illuminated from the side by light incident at low angles; high regions
are bright and low regions are dark.}
\label{fig2}
\end{figure}

Our experimental apparatus is similar to that used in
Refs. \cite{melo95,umb97,umb96} and is described in detail in
Ref.~\cite{umbths}.  A layer of granular material is placed in the
bottom of an evacuated container mounted on the armature of an
electro-mechanical shaker and sinusoidally oscillated in the vertical
direction (see Fig.~\ref{fig2}(a)).  Our granular layers consist of
spherical non-cohesive and non-magnetic particles of bronze, 316
stainless steel, titanium, or lead with as-poured densities of 5.1,
2.6, 4.7 and 6.7 g/cm$^3$ respectively, and with $0.05 < D < 0.8$ mm
and a poly-dispersity of $\pm 10\%$.  Data is obtained for $D=0.17$ mm
bronze particles except as noted.  Three different containers are used
(see Table 1); each has a polished aluminum base to prevent static
charge accumulation and clear plastic side walls and top for lighting
and visualization respectively. An accelerometer mounted on the
underside of the container measures the acceleration.  The patterns
are visualized using as a light source a ring of strobed
light-emitting diodes, which encircle the container and illuminate the
layer from the side (Fig.~\ref{fig2}(b)).  Images are acquired by a
digital CCD camera mounted on axis.

\begin{table}[t]
  \centering
  \caption{Container specifications.}
  \begin{tabular}{c c c c c} Container&Shape&Size (mm)&Base (mm)&Mass (g)\\
    \hline
    {\it C}&circular&126 (dia.)&10.2&1150\\
    {\it S1}&square&162 (side)&12.7&2170\\
    {\it S2}&square&106 (side)&12.7&1460\\
  \end{tabular}
\end{table}

The momentum transfer generated during layer-container contact is
measured by an acceleration compensated pressure sensor with a
resonant frequency of 300 kHz (PCB 112A22).  This sensor has a
circular sensing area 5.5 mm in diameter; the sensor size can be
compared with the wavelength of the pattern, which varies from 2 to 40
mm for the present measurements, and the particle size, typically
0.17 mm.  The sensor is flush mounted ($\pm 0.02$ mm) 13 mm from one
container sidewall and 36 mm from the other in the bottom of cell {\it
S2}.  To avoid high sampling frequencies, a circuit consisting of a
peak detector, a level shifter, and a sample-and-hold is used to
record the maximum pressure from each collision of the layer with the
container.


\section{Wave Onset}
\label{lpress}

Here we investigate the transition from a flat layer to standing waves
in the regime where the layer free-flight time $t_{flt}$ is less than
one period of oscillation of the container $T$.  The layer state is
characterized by local collision pressure measurements
\cite{smallsens}.  Related work is reported in Ref.~\cite{umbths} and
in Refs.~\cite{surf,mmpp}, where in the latter, global force and
optical measurements characterize the flat layer state.  For $\Gamma <
1$, the layer is always in contact with the plate and there is no
significant relative grain motion.  For $\Gamma > 1$, we identify
three distinct stages of layer motion during each cycle: free-flight
-- layer not in contact with container; impact -- layer and container
collide; contact -- layer and container in contact and moving with the
same vertical velocity.  Impact imparts relative kinetic energy to the
grains which, if large enough, enables the layer to change its
configuration.

\begin{figure}[tb]  	
\centering \includegraphics[width=0.49\textwidth]{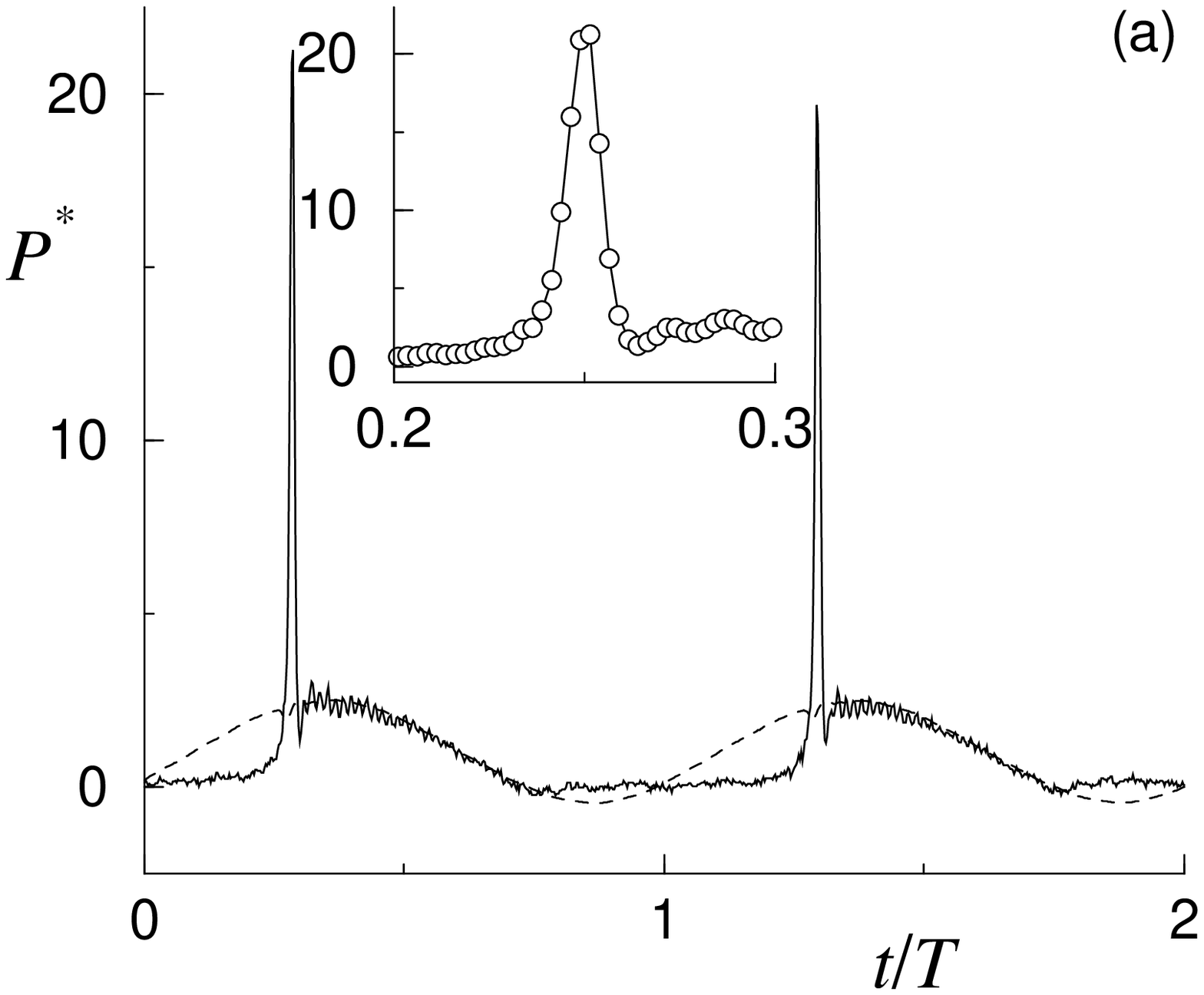}
\hfill \includegraphics[width=0.49\textwidth]{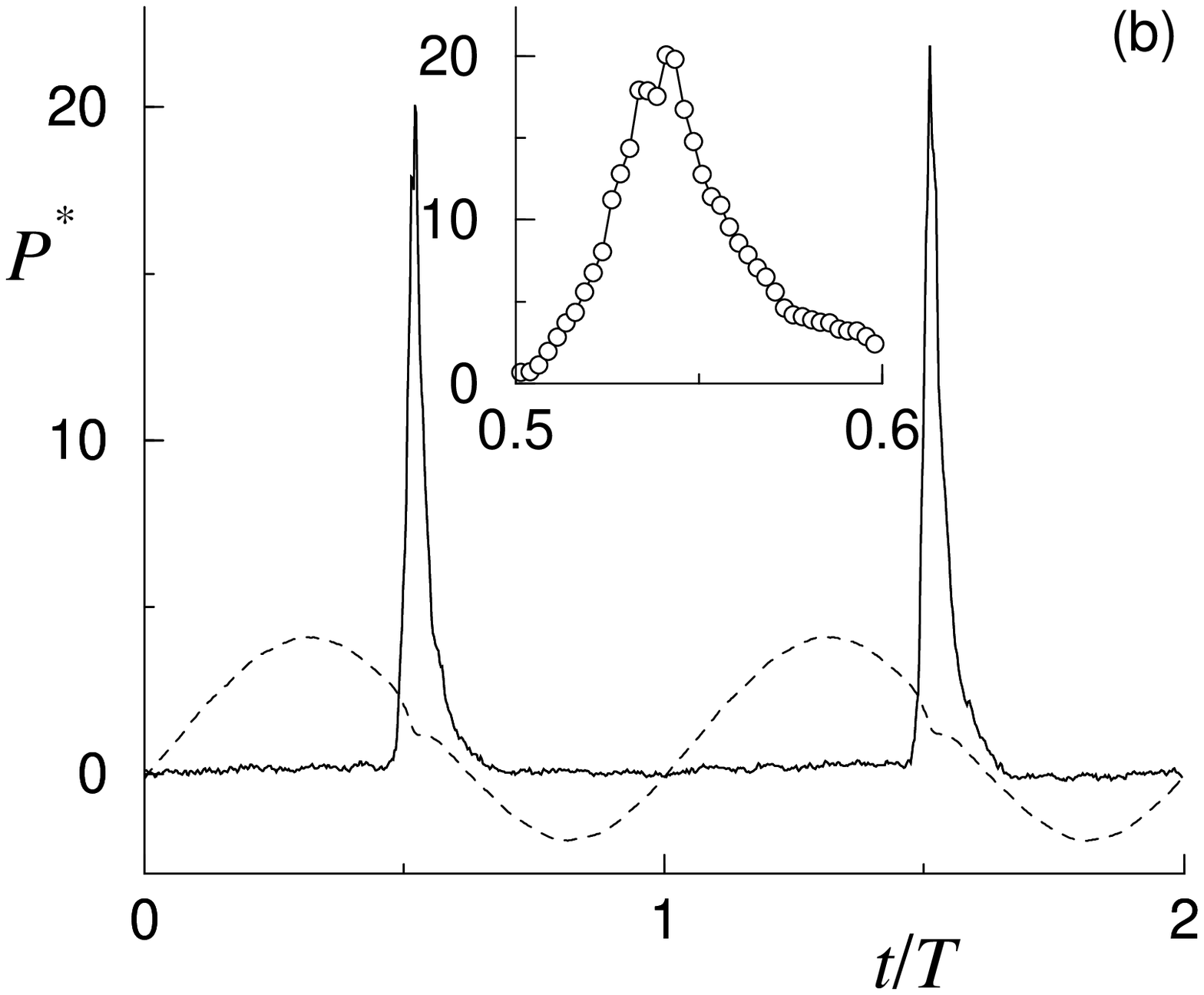}
\caption{Non-dimensionalized layer pressure for (a) $\Gamma=1.5$ (flat
layer) and (b) $\Gamma=3.1$ (waves) with $f^*=0.48$ ($f=40$ Hz) and
$N=9$. Dashed curve is a plot of $a(t)/g+1$.  The insets show expanded
portions of the collision region and illustrate the increase in
collision time associated with waves onset.}
\label{fig3}
\end{figure}

We first describe some generic features of the layer-container
pressure in the vicinity of the flat-layer/waves transition.  A time
series for the local pressure exerted by the granular layer on the
container below the onset of the waves is plotted in
Fig.~\ref{fig3}(a).  The pressure is expressed in non-dimensional
form, $P^* = P/\rho g H$, where $P$ is the pressure and $\rho$ is the
as-poured layer density.  $P^* = 0$ indicates the layer is in
free-flight, the sharp jump in $P^*$ occurs at impact, while contact
is shown by $P^* \approx a(t)/g + 1$ --- the pressure calculated using
the measured plate acceleration and assuming a solid layer attached to
the plate (dashed curve).  Several features of $P^*$ should be noted.
As the inset in Fig.~\ref{fig3}(a) emphasizes, the collision duration
is short, less than $0.02 T$, which indicates the layer is relatively
compact at the time of collision.  Furthermore, immediately following
the collision, $P^* = a(t)/g + 1$, which shows the collision is
strongly inelastic (if the layer bounced, $P^*$ would be reduced).
Finally, $P^*$ and $a(t)/g+1$ go to zero simultaneously, which
indicates the entire layer is moving upward with the container
velocity when free-flight begins.

To contrast the layer behavior above and below wave onset, a time
series of $P^*$ in the wave regime is presented in Fig.~\ref{fig3}(b).
Here the layer motion consists of just two stages, a free-flight
followed by an extended collision.  There is no longer a contact stage
as indicated by the observation that $P^*$ is always larger than
$a(t)/g+1$, and that $a(t)/g+1$ goes to zero before $P^*$ goes to
zero.  From these observations, we infer that the layer is expanded at
impact, that at no time during the collision is the entire layer
totally at rest with respect to the container, and that the collision
continues for a short time after $a(t)$ becomes less than $g$.

\begin{figure}[tb]  	
\centering
\includegraphics[width=0.49\textwidth]{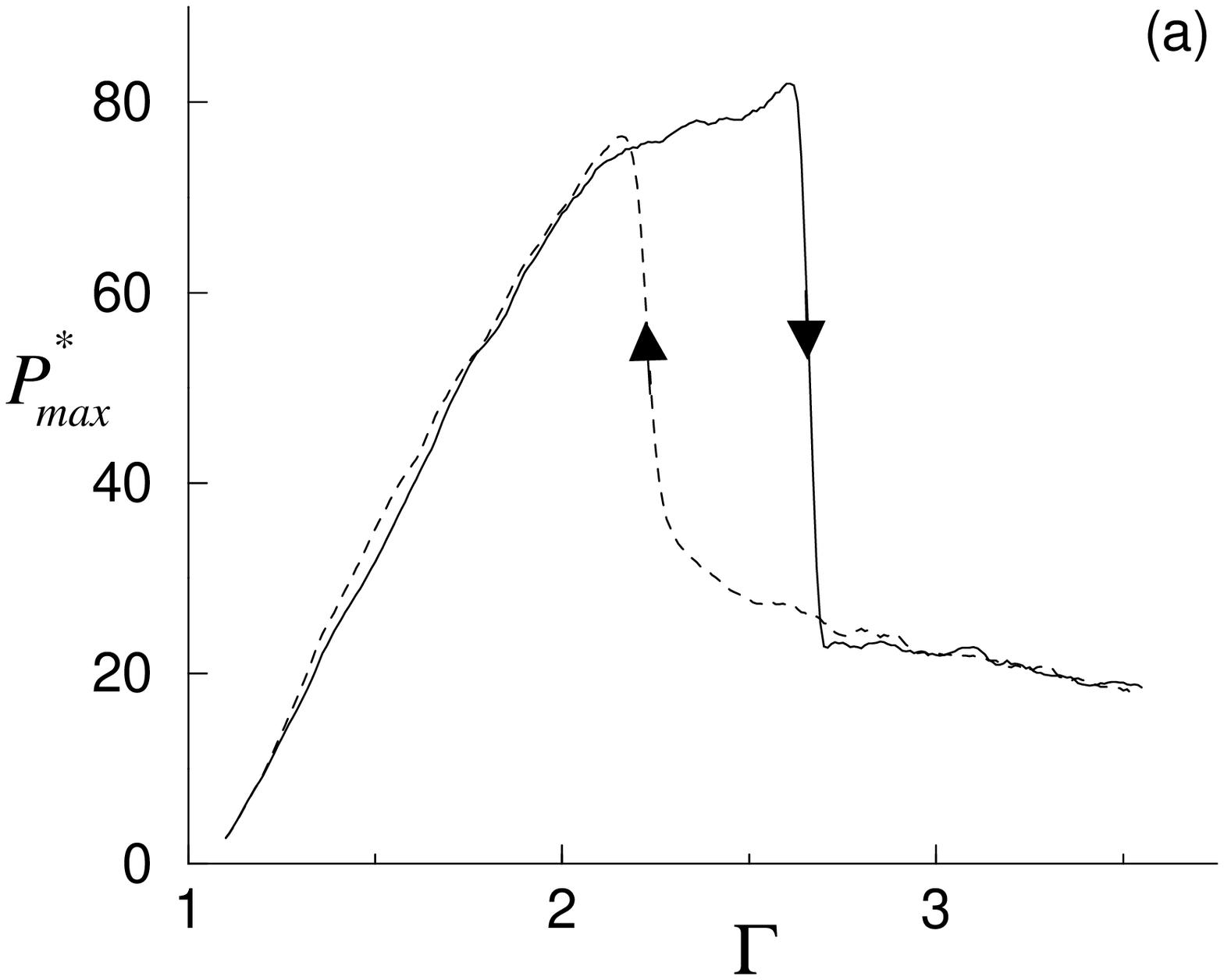}%
\hfill \includegraphics[width=0.49\textwidth]{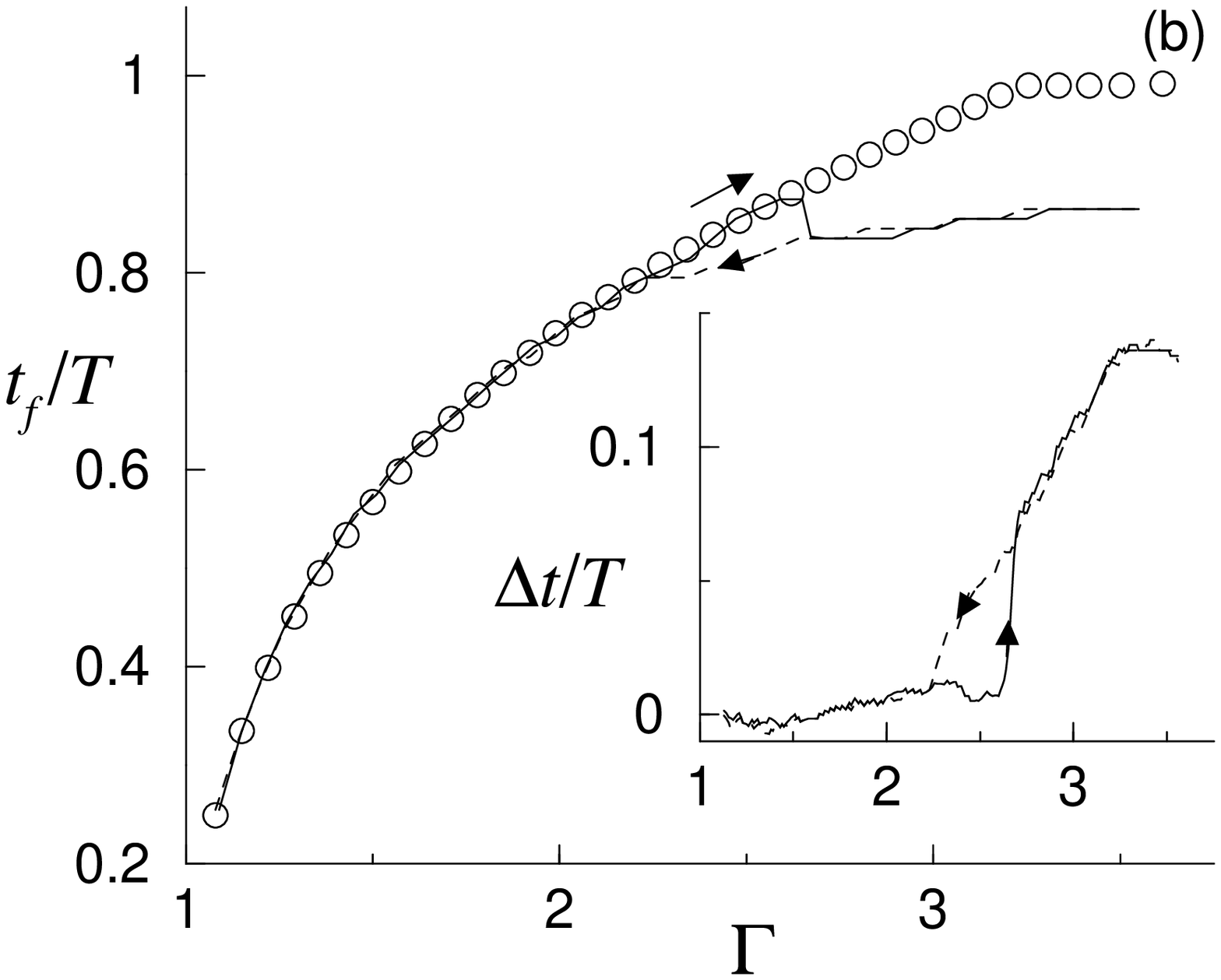}
\caption{The hysteresis in the transition from a flat layer to waves
is revealed in measurements of (a) maximum collision pressure and (b)
dimensionless flight time versus acceleration amplitude for increasing
(solid-line) and decreasing (dashed-line) $\Gamma$ ($f=26$ Hz
($f^*=0.31$), $N=8.2$).  The inset in (b) plots the difference $\Delta
t$ between the flight time calculated from the inelastic layer model
({\large $\circ$} in (b)) and the experimental flight time.
Experimental data are averaged into $0.01$ intervals in $\Gamma$ and
averaged over approximately 20 full ramps of $\Gamma$.}
\label{fig4}
\end{figure}

At the onset of waves, the maximum collision pressure per cycle,
$P^*_{max}$, decreases rapidly, as Fig.~\ref{fig4}(a) illustrates.
This decrease is associated with an increase in layer dilation (see
Fig.~\ref{fig3}).  With increasing $\Gamma$, the decrease in
$P^*_{max}$ occurs at $\Gamma=2.7$.  However, the transition is
hysteretic and the layer reverts to the flat state at $\Gamma=2.2$,
where $P^*_{max}$ increases abruptly (see Fig.~\ref{fig4}(a)).  For
the flat layer, $P^*_{max}$ increases with increasing $\Gamma$, while
in the wave state, $P^*_{max}$ decreases with increasing $\Gamma$
\cite{pnote}.  Additional measurements using steel and titanium
particles show a somewhat different behavior \cite{umbths}.  Before
the transition to waves ($\Gamma \approx 1.9$), $P^*$ in these larger
restitution coefficient materials drops from 80 to approximately 40
but then continues to increase until the onset of waves at which point
$P^*$ drops to 20 as for the bronze layers.

Figure \ref{fig4}(b) compares the layer free-flight time measured from
the experiment with the time calculated from the completely inelastic
layer model \cite{mehta,melo95}.  When the layer is flat, the measured
and calculated flight times are nearly identical.  However, when waves
arise, the measured flight time is smaller than the model predicts
(see inset in Fig.~\ref{fig4}(b)).  This discrepancy arises because
the velocity of a portion of the layer is less than the container
velocity when $a(t)$ becomes less than $g$.  These slower grains
reduce the effective layer take-off velocity, which decreases
$t_{flt}$.  For increasing $\Gamma$, there is a sudden decrease in
$t_{flt}$ when waves arise.  However, for decreasing $\Gamma$, and
unlike $P^*_{max}$, $t_{flt}$ continuously decreases until it equals
the model value at which point the layer reverts to the flat state.
The equality of measured and model values of $t_{flt}$ at this
transition implies that for waves to exist, a portion of the layer
must be dilated when the bottom of the layer leaves the container.
This necessary condition appears to be due to the strongly dissipative
nature of the granular layer, which rapidly removes kinetic energy
when layers remain in contact with the container.

\begin{figure}[tb]  	
\centering \includegraphics[height=0.43\textwidth]{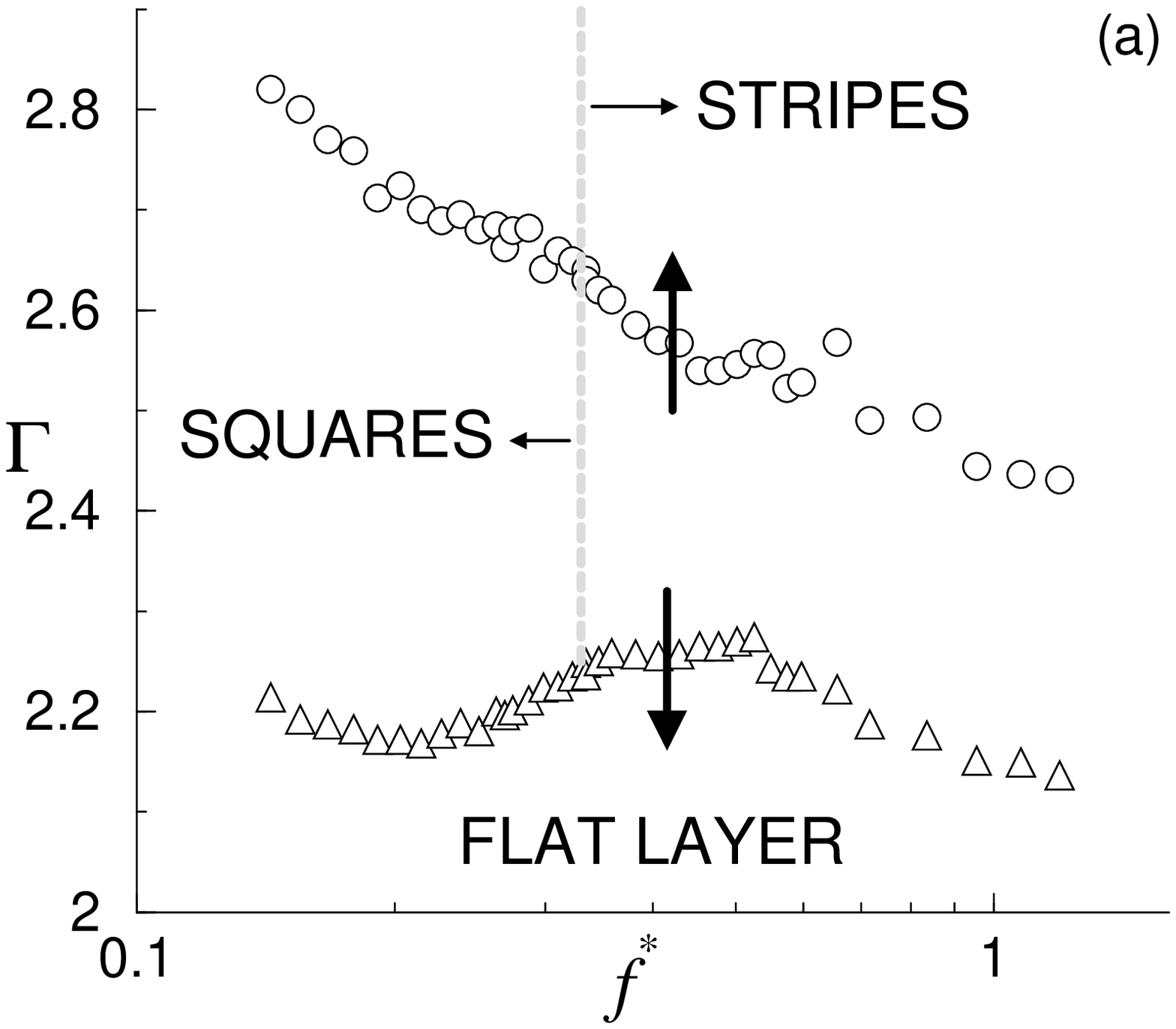}
\hfill \includegraphics[height=0.43\textwidth]{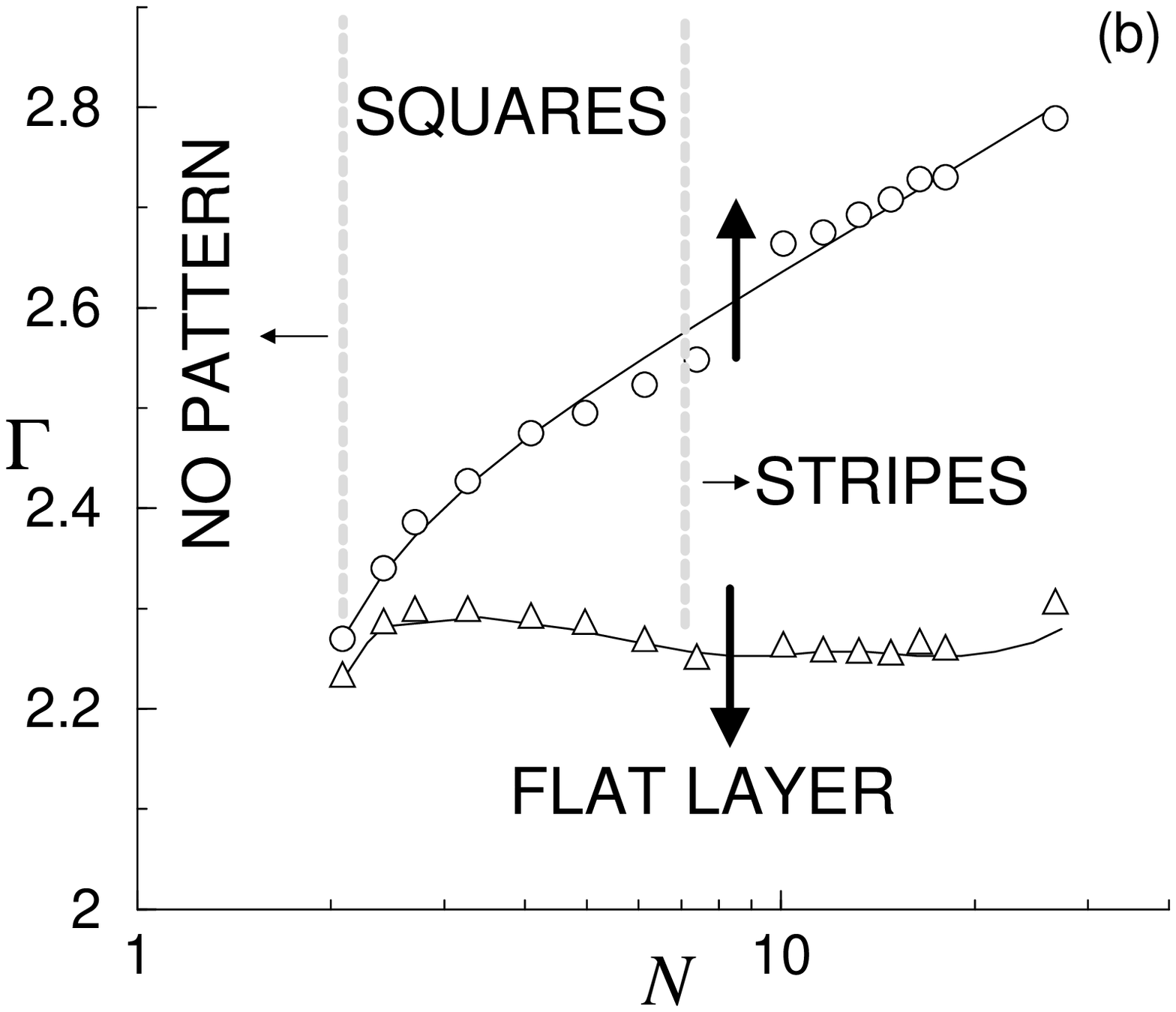}
\caption{Stability diagram showing pattern transitions with increasing
({\large $\circ$}) and decreasing ($ \triangle $) $\Gamma$ as a
function of (a) $f^*$ with dimensionless layer thickness $N=9$, and
(b) $N$ with $f^*=0.36$ ($f=30$ Hz).  Hysteresis decreases with
increasing $f^*$ and increases with increasing $N$.  In (b) for $N
\lesssim 2$ the hysteresis vanishes; as $\Gamma$ is increased in this
regime the layer dilates into a uniform low density gas-like state.}
\label{fig5}
\end{figure}

The qualitative features of the transition from a flat layer to waves,
shown in Figs. \ref{fig3} and \ref{fig4}, are the same for the entire
range of $H$, $D$, and $f$ we examine.  However, other features of the
transition are non-trivial functions of $H$ and $f$.  For instance,
the $\Gamma$ value at which the transition from a flat layer to waves
occurs, $\Gamma_c$, is plotted in Fig.~\ref{fig5}(a) as a function of
$f^*$.  For increasing $\Gamma$, $\Gamma_{c\uparrow}$ decreases
monotonically with $f^*$, while for decreasing $\Gamma$,
$\Gamma_{c\downarrow}$ is nearly constant except for a slight local
increase near $f^* \approx 0.4$ \cite{crtgam}.  The observation that
the existence of waves depends on the layer never reaching the contact
stage gives the following necessary condition for wave existence
\begin{equation}
T-t_{flt} = t_{coll} = h / v_{coll},
\label{dilation}
\end{equation}
where $h$ is the layer dilation (actual layer thickness minus $H$),
$t_{coll}$ is the collision duration, and $v_{coll}$ is the collision
velocity.  At the flat-layer/waves transition for decreasing $\Gamma$,
$\Gamma_{c\downarrow}$ is nearly constant which implies $h \propto
f^{-2}$ since $T-t_{flt}$ and $v_{coll}$ are both proportional to
$f^{-1}$. For the data in Fig.~\ref{fig5}(a), $\Gamma_{c\downarrow}
\approx 2.2$ which gives $h = 0.25 v_{coll} t_{flt}$.  Similarly,
applying Eq.~(\ref{dilation}) in the flat layer regime where
$\Gamma_{c\uparrow}$ is a function of $f$, yields $h \propto f^{-1.8}$
--- a slower decrease with increasing $f$ than for waves.

$\Gamma_c$ is also a function of $N$, as Fig.~\ref{fig5}(b) shows.
$\Gamma_{c\downarrow}$ is nearly independent of $N$, while
$\Gamma_{c\uparrow}$ is roughly proportional to $N$. The transition as
measured by the rapid drop in $P_{max}$ is no longer subcritical for
$N \lesssim 2$ --- the same $N$ value below which coherent wave
patterns disappear.  Interpreting these results in light of
Eq.~(\ref{dilation}), the increase in $\Gamma_{c\uparrow}$ with
increasing $N$ is due to the greater dissipation in the thicker layer
which reduces $h$.  The disappearance of waves for $N \leq 2$ is
likely a result of large grain velocity fluctuations that destroy
coherent layer motion.  This interpretation is also supported by
experiments with different materials \cite{umbths}: materials with a
restitution coefficient larger than that of bronze, such as stainless
steel and titanium, require deeper layers than does bronze to generate
waves. Lead, which has a smaller restitution coefficient than bronze,
requires a layer of only $N = 1.1$ for waves to exist.  Additionally,
in layers deep enough to form waves, we find that higher (lower)
restitution materials have higher (lower) values of $\Gamma_c$ for
equal $N$.

In Ref.~\cite{bizon98}, Bizon {\it et al.} note that waves in their
simulations form when the time for peak growth is greater than that
for peak decay.  From this observation, they calculate
$\Gamma_{c\downarrow} \approx 2.47$ independent of either $f$, $D$, or
$H$.  The data presented above is in reasonable agreement with their
prediction.  Their observation does not address the internal layer
state and thus can not predict the value of $\Gamma_{c\uparrow}$ and
its dependence on $f$ and $N$.  With results from simulations, it will
be possible to measure $h$ and thus check the validity of the criteria
for wave onset proposed in Eq.~\ref{dilation}.

\section{Square/Stripe Transition}
\label{sqsttrans}

\begin{figure}[t]  	
  \centering \includegraphics[width=0.57\textwidth]{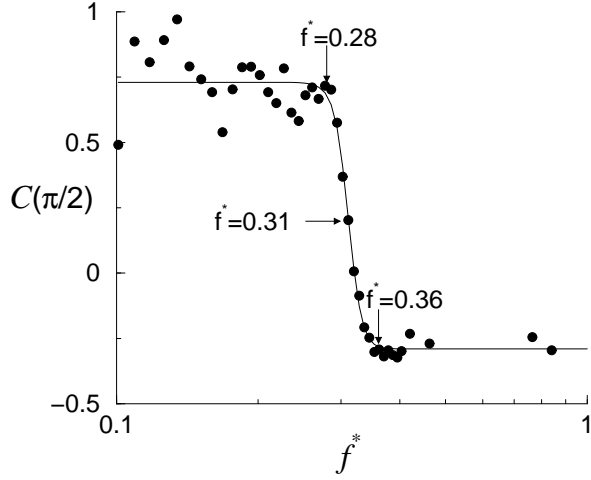} \\
  \caption{The transition from square patterns at low frequencies to
  stripe patterns at high frequencies is indicated by the decrease in
  the autocorrelation function of radially averaged power spectral
  density at $\theta = \pi/2$ ($\Gamma=2.5$, $N=4$, {\it S1}). The
  solid line is a fit to $a + b\tanh[c(f^*-f^*_{ss})]$ with $f^*_{ss}
  = 0.31$. The labeled $f^*$ values correspond to the images in
  Fig.~\ref{fig1}.}  \label{fig6}
\end{figure}

The wave patterns at the primary bifurcation are squares for low $f^*$
and stripes for high $f^*$, while for intermediate $f^*$, both
patterns are observed simultaneously (see Fig.~\ref{fig1}).  To
quantify the square/stripe transition, we divide the pattern images
into $3 \lambda \times 3 \lambda$ regions, calculate the spatial power
spectra, radially average the power within one full width at half
maximum of the dominant wavenumber, and subtract the mean to obtain
$I(\theta)$.  The autocorrelation function \mbox{$C(\pi/2) = {\langle
I(\theta) I(\theta+\pi/2) \rangle}_\theta/ {\langle I^2(\theta)
\rangle}_\theta$} is then used to characterize the pattern.  For
perfect squares $C(\pi/2) = 1$, whereas for an image of perfect
stripes $C(\pi/2) = 0$.  Differences between the measured and ideal
values ({\it i.e.}, $C(\pi/2) < 0$) are due to the finite angular
width of the spectral peak.  Figure \ref{fig6} is a plot of $C(\pi/2)$
versus $f^*$, which shows the transition from squares to stripes
occurs in a narrow frequency range about the transition frequency
$f^*_{ss} = 0.31$, with $C(\pi/2)$ nearly constant above and below
$f^*_{ss}$.  The value of $f^*_{ss}$ is not sensitive to details of
the method: plotting $I(\theta_{max}\pm\pi/2)$, where $\theta_{max}$
is the azimuthal location of the maximum value of $I(\theta)$, as well
as varying the sub-image size, yield the same result.

\begin{figure}[b]  	
\centerline{
\includegraphics[width=0.485\textwidth]{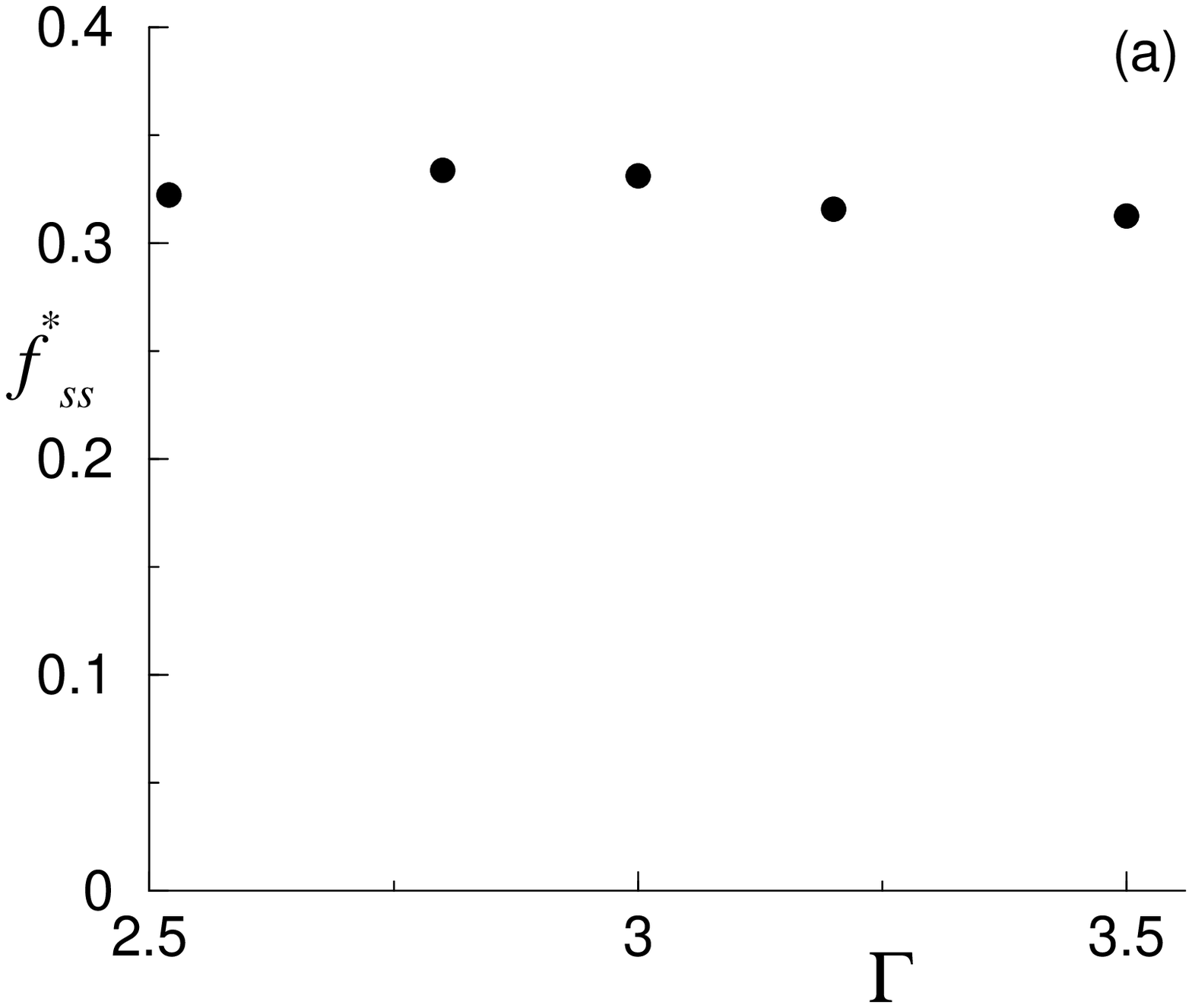} \hfill
\includegraphics[width=0.485\textwidth]{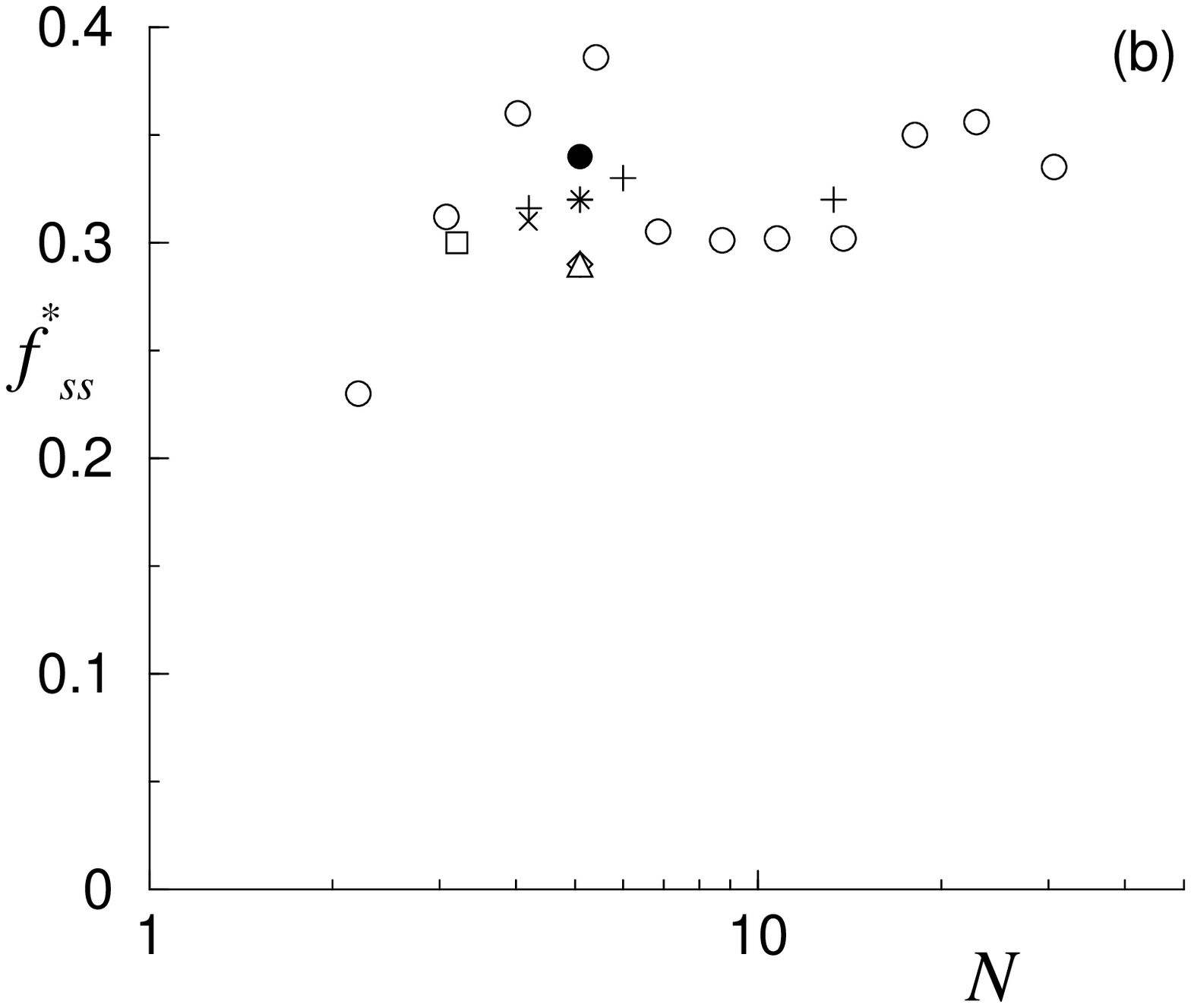}}
\caption{(a) Square/stripe transition frequency as a function of
$\Gamma$ for $N = 13$ (container {\it S1}). (b) Square/stripe
transition frequency as a function of $N$: {\large{{\large $\circ$}}}
($\Gamma = 3.0$, 0.17 mm, $C$), $+$ ($\Gamma=2.5$, 0.17 mm, {\it S1}),
$\times$ ($\Gamma=3.1$, 0.66 mm, {\it S1}), $\square$ ($\Gamma=2.5$, 0.46
mm lead, $C$), $\ast$ ($\Gamma=3.0$, 0.07 mm, $C$), $\bullet$
($\Gamma=3.0$, 0.78 mm, $C$), {\large $\diamond$} ($\Gamma=2.8$, 0.17
mm steel, $C$), $\triangle$ ($\Gamma=2.8$, 0.17 mm titanium, $C$).}
\label{fig7}
\end{figure}

Figure \ref{fig7}(a) is a plot of $f^*_{ss}$ as a function of $\Gamma$
for fixed $N$.  There is little change in $f^*_{ss}$ as $\Gamma$ is
increased from near wave onset to close to the transition to hexagons
\cite{melo95}.  Figure \ref{fig7}(b) plots $f^*_{ss}$ against $N$.
Although $N$ varies by more than an order of magnitude, the
non-dimensionalized square/stripe transition frequency remains nearly
fixed.  Included in Fig.~\ref{fig7}(b) are additional measurements
that indicate $f^*_{ss}$ also does not depend strongly on $D$,
material, or container size or shape.  

Bizon {\it et al.} have proposed that the square/stripe transition
occurs when the distance that the layer falls from its maximum height
to the container is equal to the layer depth: for fall distances
larger (smaller) than $H$, squares (stripes) are the preferred pattern
\cite{bizon98}.  Using a simple model that considers the layer as a
totally inelastic object with no internal modes \cite{mehta,melo95},
they calculate $f^*_{ss} = 1/\sqrt{8} \approx 0.35$.  The data
presented here gives an average value of $f^*_{ss} \approx 0.33$ over
a large range of experimental parameters.  The preceding measurements
as well as the criterion of Bizon {\it et al.}  suggest that the
square/stripe transition is primarily a function of the layer depth
$H$.  Layers with differing $N$ and $D$ but equal $H$ make the
transition from squares to stripes at the same $f$ (except for layers
with $N \sim O(1)$ where the pattern weakens due to large relative
particle velocities).  This result implies that the layer acts as a
continuum, {\it i.e.,} the size of the particles does not affect the
bulk behavior although the discrete nature does.  In contrast, the
dependence of the critical acceleration amplitude for waves on $N$ and
$f$, where scaling with $H$ fails (Fig.~\ref{fig5}(b)), suggests that
the physics determining subcriticality and square/stripe pattern
selection are unrelated.

\section{Wavelength Scaling}
\label{wlscaling}

The dispersion relation for surface waves in inviscid fluids in the
absence of surface tension and in the shallow water limit ($\lambda >
H$) is $\lambda/H = f^{-1} \sqrt{g/H}$.  When plotted in this form,
which uses $H$ and $\sqrt{H/g}$ as the characteristic length and time
scales respectively, the dispersion relations for different depth
layers lie on a single curve.  In granular layers, however, there are
two natural length scales -- layer depth and grain diameter.  We show
in this section that below a critical frequency, wavelength scaling is
governed by $H$ as is the case for fluids, while above this frequency,
scaling with $H$ fails.  In the next section (Sec.~\ref{gmt}), we will
show that the scaling breakdown is associated with a transition in the
horizontal grain mobility.

\begin{figure}[b]  	
\centerline{\includegraphics[height=0.42\textwidth]{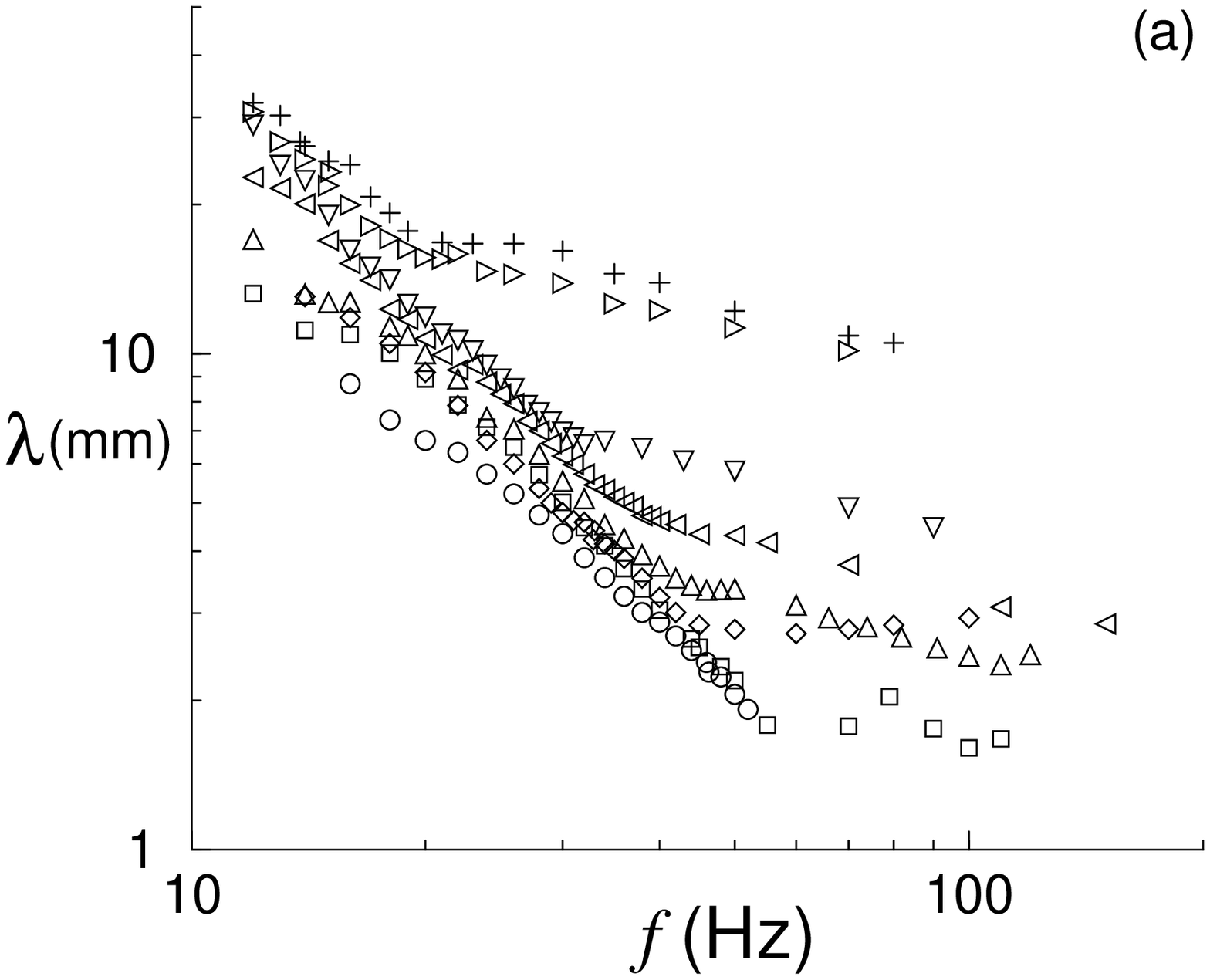} \hfill
\includegraphics[height=0.42\textwidth]{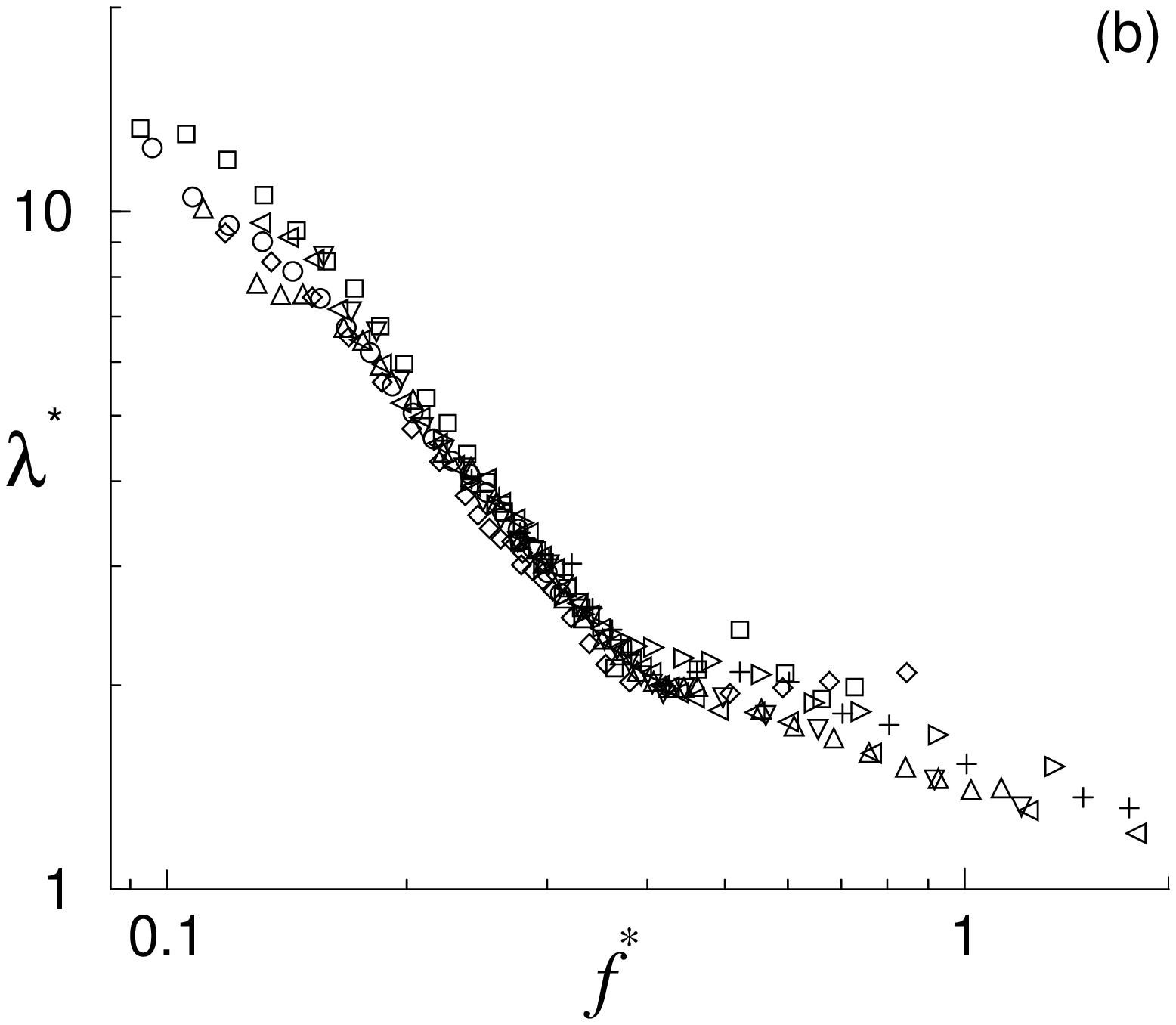}}
\caption{Wavelength as a function of frequency for layers of constant
dimensionless depth $N=H/D$ and varying particle diameter $D$.  (a)
$\lambda$ versus $f$ for $0.07$ mm ({\large $\circ$}), $0.08$
mm ($\square$), $0.14$ mm ({\large $\diamond$}), $0.17$ mm
($\triangle$), $0.23$ mm ({\large $\triangleleft$}), $0.33$ mm
({\footnotesize $\bigtriangledown$}), $0.66$ mm ({\large
$\triangleright$}), and $0.78$ mm ($+$) ($\Gamma = 3.0$, $N=5$,
{\it S1}). (b) Same data plotted with dimensionless variables using $H$ as
the fundamental length scale.  Near $f^*=0.4$ in (b), there is a kink
in the dispersion curve, which moves to higher $f^*$ as $N$ is
increased.}
\label{fig8}
\end{figure}

Figure \ref{fig8}(a) presents results for the wavelength as a function
of frequency for layers of bronze particles with $N=5$, $\Gamma=3.0$,
and a range of particle sizes ($0.07 < D < 0.8$ mm).  The wavelength
decreases with increasing $f$ and increases with increasing $D$.  When
$H$ is used to non-dimensionalize the wavelength $\lambda^* = \lambda
/ H$ and the frequency $f^* = f \sqrt{H/g}$, the data collapse onto a
single curve (Fig. \ref{fig8}(b)).  A well-defined kink in the
dispersion curve occurs near $f^*=0.4$, above which $\lambda^*$
decreases more slowly with increasing $f^*$.  A similar slowing has
been observed in other experiments but only for one \cite{metcalf} and
two \cite{bizon98} disjoint values of $D$.

\begin{figure}[t]
\centerline{\includegraphics[height=0.42\textwidth]{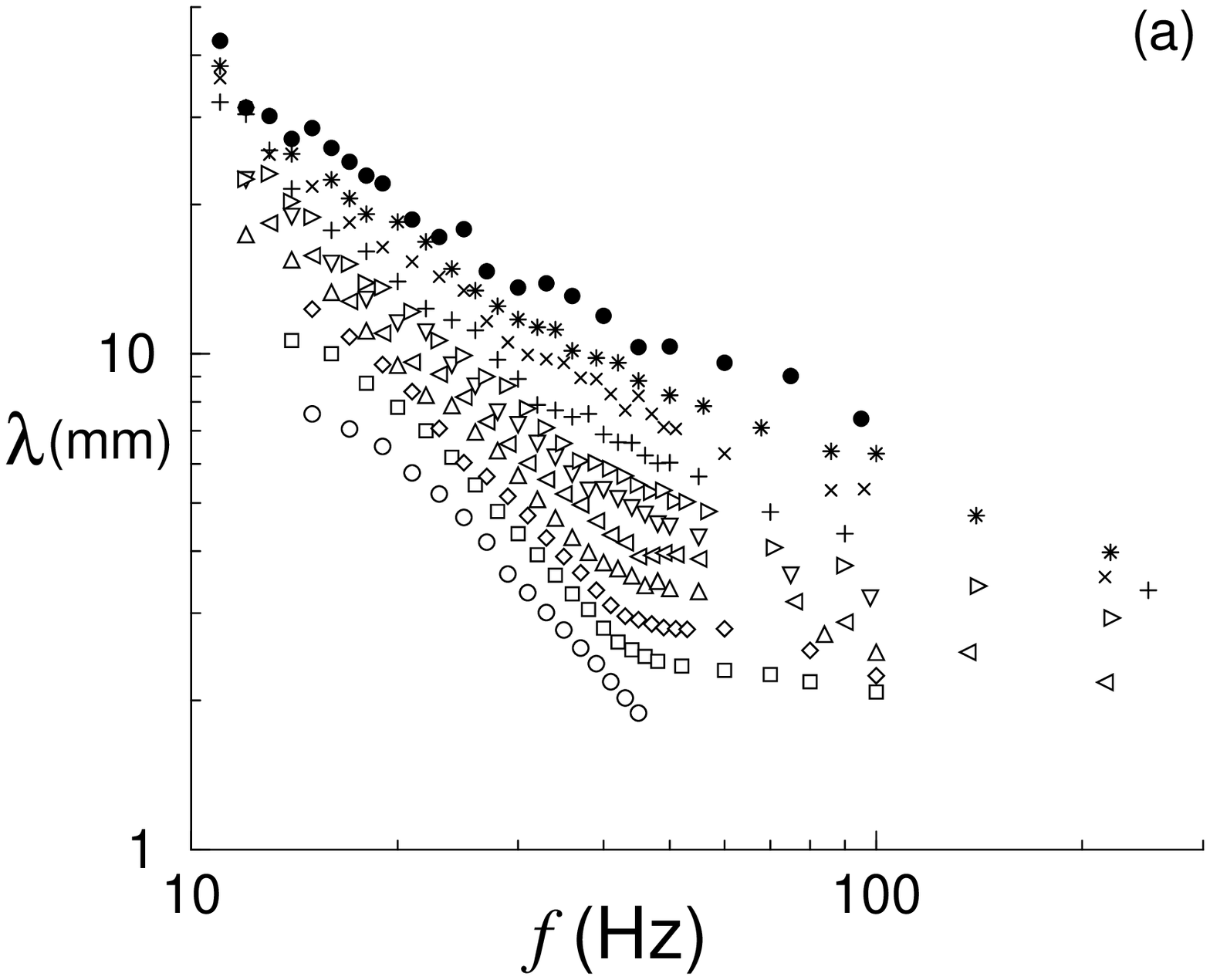}
   \hfill \includegraphics[height=0.42\textwidth]{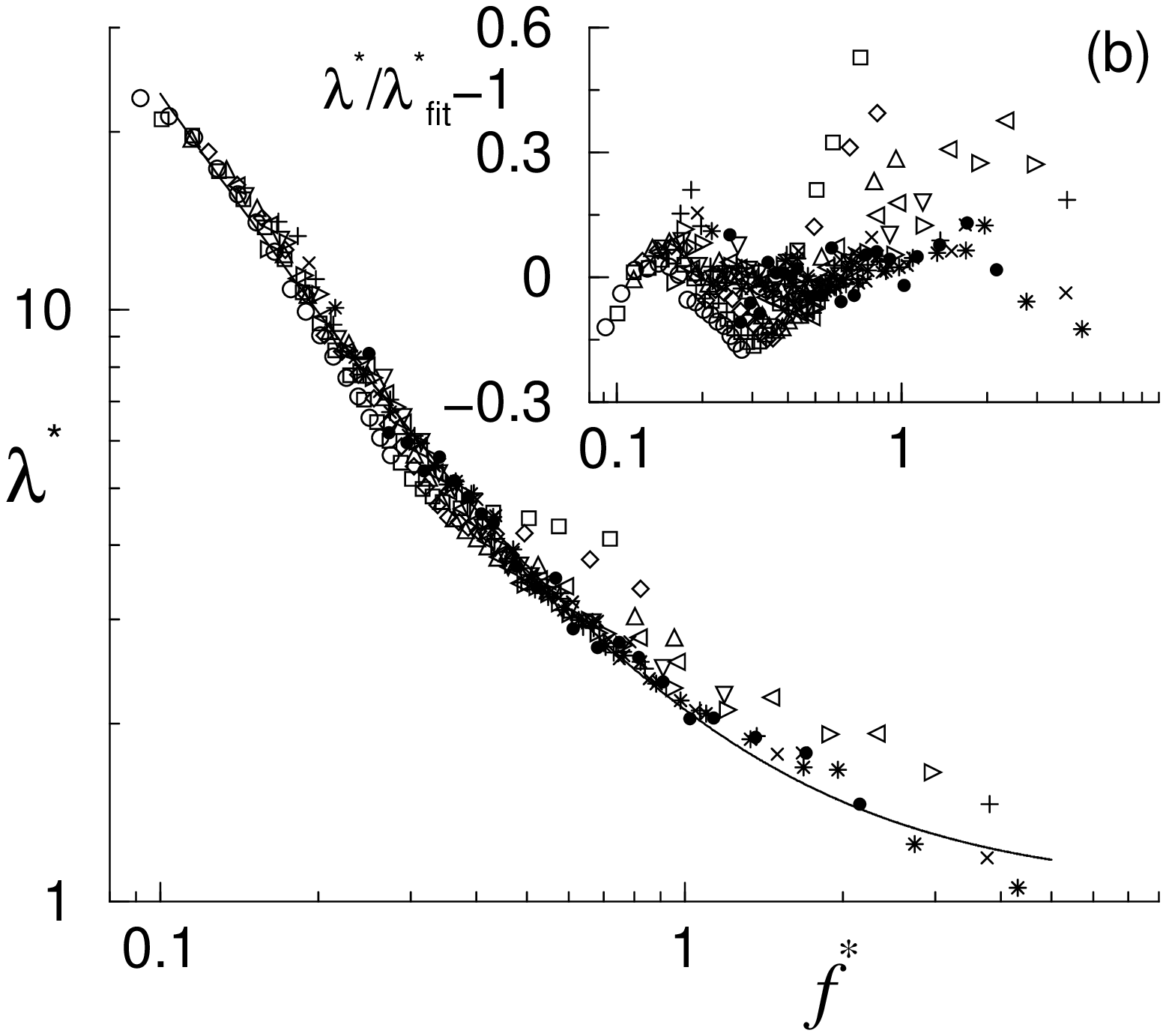}}
   \caption{Dispersion relation for layers of constant particle
   diameter and varying dimensionless depth, $N=H/D$: (a) Wavelength
   as a function of frequency: $N=2.2$ ({\large $\circ$}), $N=3.1$
   ($\square$), $N=4.0$ ({\large $\diamond$}), $N=5.4$ ($\triangle$),
   ({\large $\triangleleft$}) $N=6.8$, $N=8.7$ ({\footnotesize
   $\bigtriangledown$}), $N=10.7$ ({\large $\triangleright$}),
   $N=13.8$ (+), $N=18.1$ ($\times$), $N=22.8$ ($\ast$), and $N=30.7$
   ($\bullet$) ($D=0.17$ mm, $\Gamma=3.0$, container $C$). (b) Same
   data plotted with dimensionless variables.  The dimensionless data
   in (b) are fit to Eq.~\ref{dispeqn}; the inset shows the residuals.
   As $N$ increases, the $f^*$ range over which data collapse occurs
   increases.}
\label{fig9}
\end{figure}

Because $N$ is constant for the data presented in Fig.~\ref{fig8}, it
is unclear whether $H$ or $D$ is the length scale governing the
collapse since $H = N D$. To test whether the data collapse obtained
in Fig.~\ref{fig8}(b) is due to scaling with $H$, we plot in
Fig.~\ref{fig9} the wavelength versus frequency with fixed \mbox{$D =
0.17$ mm} but with $N$ varying from 2 to 31.  $\lambda$ increases with
increasing $H$ and again the unscaled data collapse when $\lambda^*$
is plotted as a function of $f^*$.  Similar results were obtained by
Bizon {\it et al.} but only for two values of $N$ ($N=2.7,5.4$)
\cite{bizon98}.

The rescaled data in Fig.~\ref{fig9}(b) fit the functional form
$\lambda^* = a + b {f^*}^c$ with the best fit given by 
\begin{equation}
\lambda^* = 1.0 + 1.1 {f^*}^{-1.32 \pm 0.03}.
\label{dispeqn}
\end{equation}
The residuals from the fit to the rescaled dispersion data in
Fig.~\ref{fig9}(b) (see inset) for $f^* \geq 0.4$ exhibit a marked
decrease in the quality of the data collapse, which is a function of
$H$ and which is characterized by a slowing in the decrease of
$\lambda^*$ with increasing $f^*$.  This slowing is more pronounced in
small $N$ layers.  We will return to this observation in the next
section.

Three points should be noted concerning the dispersion relation data.
First, although the granular data collapse below the kink in the
dispersion curves when scaled with $H$ as for fluid surface waves, the
dispersion relations for these two media are different.  Second, the
exponent in Eq.~\ref{dispeqn} is not equal to $-2$ as was found in our
earlier work on glass beads in air \cite{melo94}, by Metcalf {\it et
al.}  \cite{metcalf} for glass beads in an evacuated container, and by
Cl{\'e}ment {\it et al.} in a two-dimensional layer of aluminum balls
\cite{clement}.  Possible explanations for the larger value of
exponent found in the other studies include the influence of air
viscosity, static charging in non-conducting grains, small aspect
ratios, and drag associated with the side walls.  Finally, the fit
predicts a minimum wavelength equal to $H$ for large $f$.  The
$\lambda^*=1$ limit is clearly unattainable in small $N$ layers due to
the finite size of the layer's constituent grains; thus, scaling with
$H$ will fail for small $N$.  However, as we will show next, breakdown
of scaling normally occurs before this limit is reached due to a
reduction in the grain mobility.


\section{Grain Mobility Transition}
\label{gmt}
 
As discussed in the previous two sections, the layer depth is the
characteristic length scale which determines the scaling properties of
the square/stripe transition and the wavelength.  However, as
mentioned in Secs.~\ref{intro} and \ref{wlscaling}, the grain diameter
$D$ is also expected to influence the layer response.  Consider the
initial velocity required to raise a grain a distance $D$ in the
presence of gravity, $v_D = \sqrt{2 g D}$.  If the grain is embedded
in a plane of identical particles, horizontal motion will only be
possible if $v_D > \sqrt{2 g D}$; for lower velocities, the grain will
be stuck.  Assuming further that the fluctuation velocity of a typical
grain with respect to its neighbors is proportional to the peak
container velocity $2 \pi A f$, a relevant dimensionless parameter
characterizing grain mobility is $\tilde{\mathrm{v}} = 2 \pi A
f/\sqrt{Dg}$.  Section \ref{gm_evid} presents evidence of a
qualitative change in the layer response at a particular value of
$\tilde{\mathrm{v}}$ which we call the grain mobility transition,
$\tilde{\mathrm{v}}_{gm} \approx 2.5$, for large $\Gamma$ or small
$N$.  Section \ref{gmdat} then shows that this change appears related
to the loss of horizontal grain mobility for $\tilde{\mathrm{v}} <
\tilde{\mathrm{v}}_{gm}$.

\subsection{Changes in Layer Response}
\label{gm_evid}

\begin{figure}[tb]  	
  \centering \includegraphics[width=0.57\textwidth]{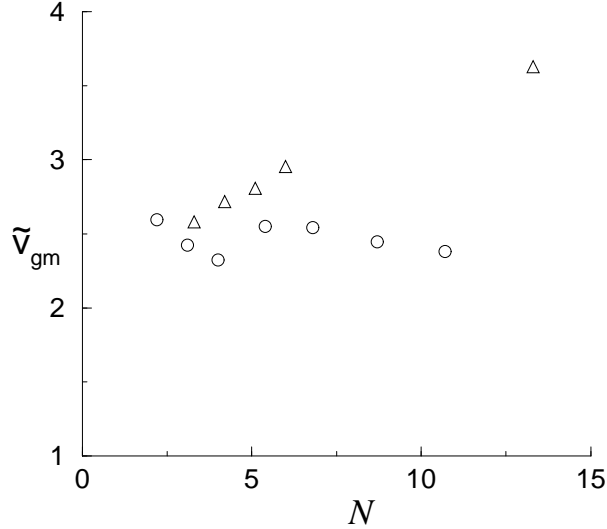}
  \caption{Dimensionless container velocity at which $\lambda^*$
  begins to decreases slower than Eq.~\ref{dispeqn}, as a function of
  $N$ for the data in Fig.~\ref{fig9}(b) with $\Gamma=3.0$ ({\large $\circ$})
  and for additional data \cite{umbths} with identical particles but
  with $\Gamma=2.5$ ($\triangle$).  For $\Gamma=3.0$,
  $\tilde{\mathrm{v}}_{gm}$ is nearly constant, while for
  $\Gamma=2.5$, $\tilde{\mathrm{v}}_{gm}$ increases gradually with
  increasing $N$.}  \label{fig10}
\end{figure}

The value of $f^*$ below which data collapse occurs for constant $D$
and varying $N$ is marked by a kink in the dispersion curves (see
Fig.~\ref{fig9}(b)) and is an increasing function of $N$.  The data in
Fig.~\ref{fig8}(b) for constant $N$ and varying $D$ show a similar
kink which is, in contrast to the data for constant $D$, independent
of $H$ for fixed $N$.  To characterize the location of the kink,
Fig.~\ref{fig10} presents a plot of $\tilde{\mathrm{v}}$ at the kink
({\it e.g.}  $\tilde{\mathrm{v}}_{gm}$) as a function of $N$ for
$\Gamma=3.0$ and $\Gamma=2.5$ \cite{minN}. For $\Gamma=3.0$,
$\tilde{\mathrm{v}}_{gm} \approx 2.5$ and shows no systematic
dependence on $N$.  This is nearly the same as
$\tilde{\mathrm{v}}_{gm}=2.6$, the value found for varying $D$,
constant $N=5$ and $\Gamma=3.0$ (see Fig.~\ref{fig8}). For
$\Gamma=2.5$, $\tilde{\mathrm{v}}_{gm}$ is slightly larger and is a
slowly (slower than $\sqrt{N}$) increasing function of $N$.

\begin{figure}[t]
\centering \includegraphics[width=0.57\textwidth]{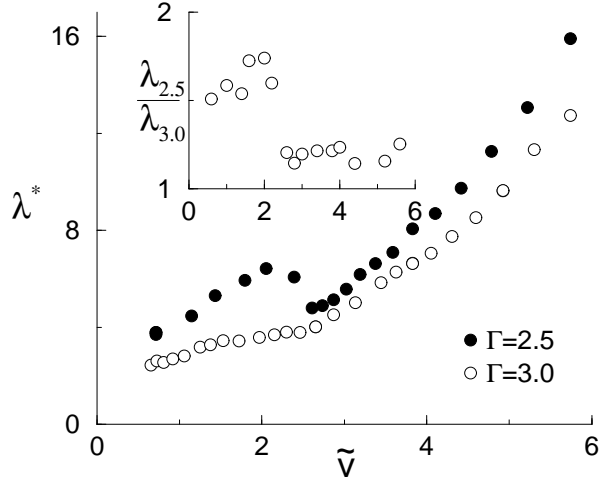}
  \caption{$\lambda^* = \lambda/H$ versus $\tilde{\mathrm{v}} = 2 \pi
  A f/\sqrt{Dg}$ for $\Gamma = 2.5$ ($\bullet$) and $\Gamma=3.0$
  ({\large $\circ$}) (0.46 mm lead, $N=5$, container $C$).  Inset:
  The $\lambda$ ratio jumps for $\tilde{\mathrm{v}} < 2.6$, which
  indicates the grain mobility transition.}  \label{fig11}
\end{figure}

To summarize, for $\mathrm{v} > \mathrm{v}_{gm}$, $\lambda^*$ is
independent of $H$ and scales with $f^*$, while for $\mathrm{v} <
\mathrm{v}_{gm}$, $\lambda^*$ scaling with $H$ fails.  This breakdown
appears to be related to a transition in the layer response when
$\tilde{\mathrm{v}}$ is small.  Observations at $\tilde{\mathrm{v}}
\approx 0.5$ using relatively large grains show that there is little
if any horizontal particle motion at the surface.  In contrast, for
$\tilde{\mathrm{v}} > \tilde{\mathrm{v}}_{gm}$, grains slosh back and
forth -- in thinner layers ($N\approx 4$) the sloshing is so vigorous
that the layer depth goes to zero in the pattern minima.  To
illustrate the difference between these two wave regimes,
Fig.~\ref{fig11} presents a plot of $\lambda^*$ versus
$\tilde{\mathrm{v}}$ for $N=5$ and $\Gamma=2.5$ and $\Gamma=3.0$.  The
wavelength is approximately 10 percent larger for $\Gamma=2.5$ than
for $\Gamma = 3.0$ when $\tilde{\mathrm{v}} > \tilde{\mathrm{v}}_{gm}
\approx 2.6$, but for $\tilde{\mathrm{v}} < \tilde{\mathrm{v}}_{gm}$,
$\lambda^*$ for $\Gamma=2.5$ jumps to 1.5 times $\lambda^*$ for
$\Gamma=3.0$.  If $\lambda^*$ is instead plotted versus $f^*$, the
change in behavior is even more evident: for $f^*$ below the
transition $\lambda^*$ increases with increasing $\Gamma$, while for
$f^*$ above, $\lambda^*$ decreases with increasing $\Gamma$.

\begin{figure}[t]  	
\centering \includegraphics[width=0.57\textwidth]{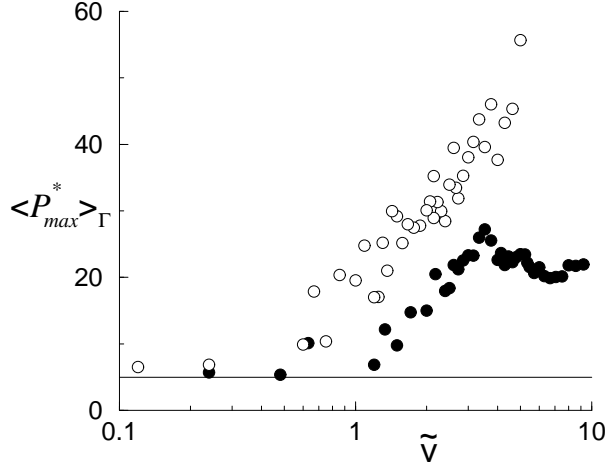} \hfill
    \caption{$\Gamma$-averaged collision pressure versus container
    velocity below wave onset for $1.2 < \Gamma < 1.8$ ({\large
    $\circ$}) and above wave onset for $2.8 < \Gamma < 3.2$
    ($\bullet$) ($N=9$). For $\tilde{\mathrm{v}} > 3$,
    ${\langle P^*_{max} \rangle}_{\Gamma}$ in the wave regime is
    constant, while for $\tilde{\mathrm{v}} < 3$, ${\langle P^*_{max}
    \rangle}_{\Gamma}$ for waves and flat-layer show similar behavior.
    The line at $\langle P^*_{max} \rangle_{\Gamma} = 5$ represents
    the calculated minimum value for $P^*_{max}$ at $\Gamma = 3.0$.}
\label{fig12}
\end{figure}

In addition to the breakdown in $\lambda$ scaling with $H$, a
qualitative change in the collision pressure is also observed for
$\tilde{\mathrm{v}} \approx 3$.  The $\tilde{\mathrm{v}}$ dependence
of the $\Gamma$ averaged collision pressure ${\langle P^*_{max}
\rangle}_{\Gamma} = \int^{\Gamma_{max}}_{\Gamma_{min}} P^*(\Gamma)
{d\Gamma}$ is plotted in Fig.~\ref{fig12} for both the flat-layer and
wave regimes.  For the flat layer, ${\langle P^*_{max}
\rangle}_{\Gamma}$ increases continuously with increasing
$\mathrm{v}$.  However, in the wave regime ${\langle P^*_{max}
\rangle}_{\Gamma}$ is nearly constant for $\tilde{\mathrm{v}} > 3$ but
decreases in a similar fashion to the flat layer for
$\tilde{\mathrm{v}} < 3$ \cite{pconst}.  The latter result suggests
that the flat layer and wave states are similar for
$\tilde{\mathrm{v}} < 3$ as would be expected below the grain mobility
transition.

Finally, as Sec.~\ref{sqsttrans} describes, the frequency of the
square stripe transition is given by $f_{ss} = 0.33 \sqrt{g/H}$ and
shows no systematic dependence on the grain size.  However, in thin
layers when $\Gamma \sqrt{N} < 2 \pi \tilde{\mathrm{v}}_{gm}
f^*_{ss}$, the frequency associated with the grain mobility transition
$f_{gm}$ is less than the frequency of the square/stripe transition.
In this case and for $\tilde{\mathrm{v}} \lesssim
\tilde{\mathrm{v}}_{gm}$, the resulting wave patterns are noticably
more tenuous and have significantly shorter spatial correlation
lengths (on the order of $\lambda$) than do the corresponding patterns
for $f_{ss} > f_{gm}$.

\subsection{Loss of Horizontal Grain Motion}
\label{gmdat}

In this section, the breakdown in $\lambda$ scaling with $H$ and the
decrease in the collision pressure are identified with a transition in
the horizontal grain mobility.  We present data suggesting that for
$\tilde{\mathrm{v}} > \tilde{\mathrm{v}}_{gm}$, the local layer height
is changed by lateral grain motion, while for $\tilde{\mathrm{v}} <
\tilde{\mathrm{v}}_{gm}$, grains are essentially immobile
\cite{thsref}.  In Ref.~\cite{mmpp}, Mujica and Melo similarly propose
that waves at low container velocities result from bending of the
layer and not from horizontal grain transfer.  Also, in our direct
visual observations we observe significant horizontal particle motion
for waves at large $\tilde{\mathrm{v}}$ but no noticeable horizontal
grain motion for waves at small $\tilde{\mathrm{v}}$.  In general for
waves at both high and low $\tilde{\mathrm{v}}$, patterns persist when
the container is rapidly brought to rest.  When the stationary
container is lightly and repeatedly tapped, the patterns at low
$\tilde{\mathrm{v}}$ slowly disappear without any apparent grain
motion on the layer surface, whereas the high $\tilde{\mathrm{v}}$
patterns exhibit significant grain rearrangement.

\begin{figure}[t]  	
  \centering
  \includegraphics[height=0.4\textwidth]{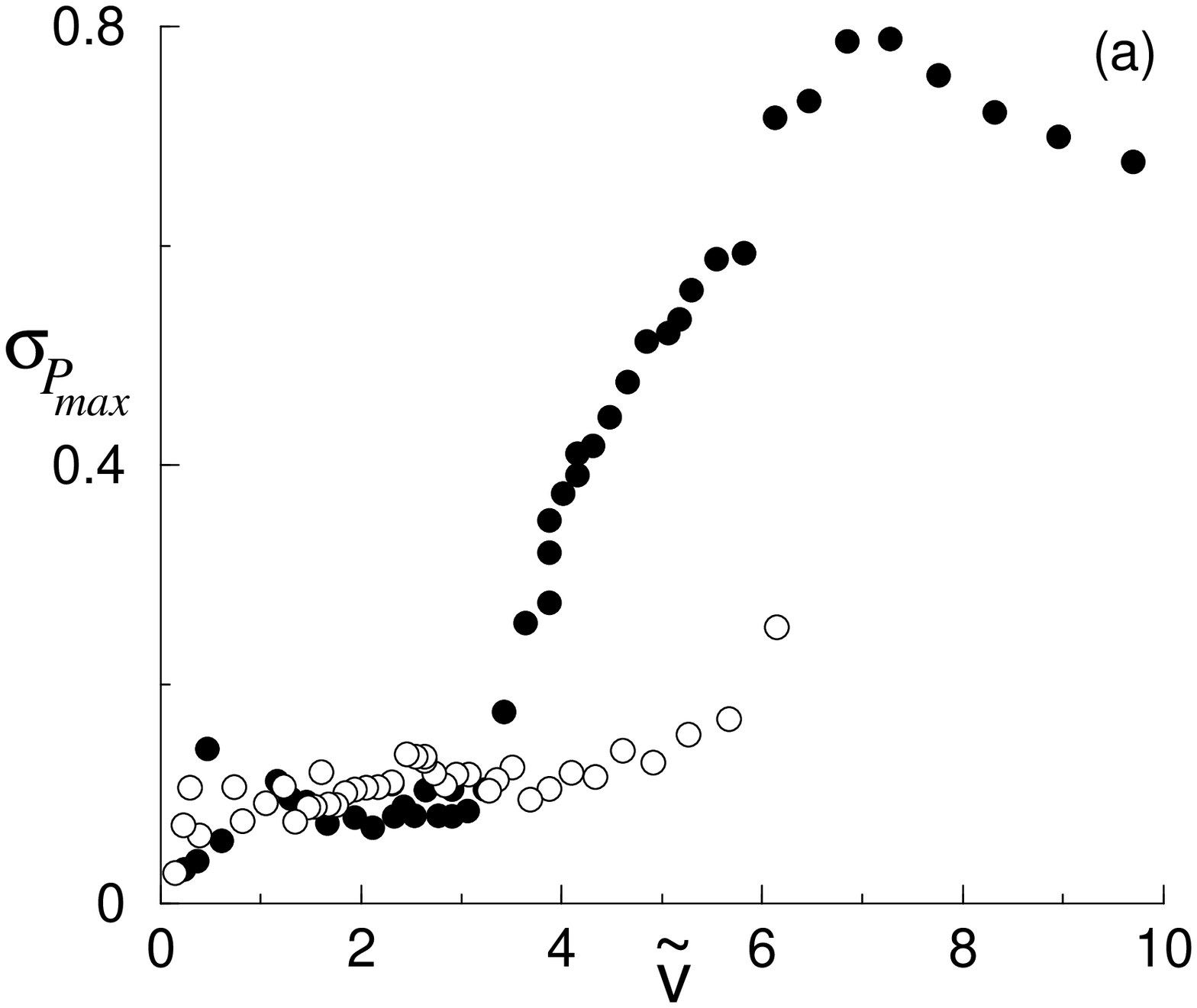}
  \includegraphics[height=0.4\textwidth]{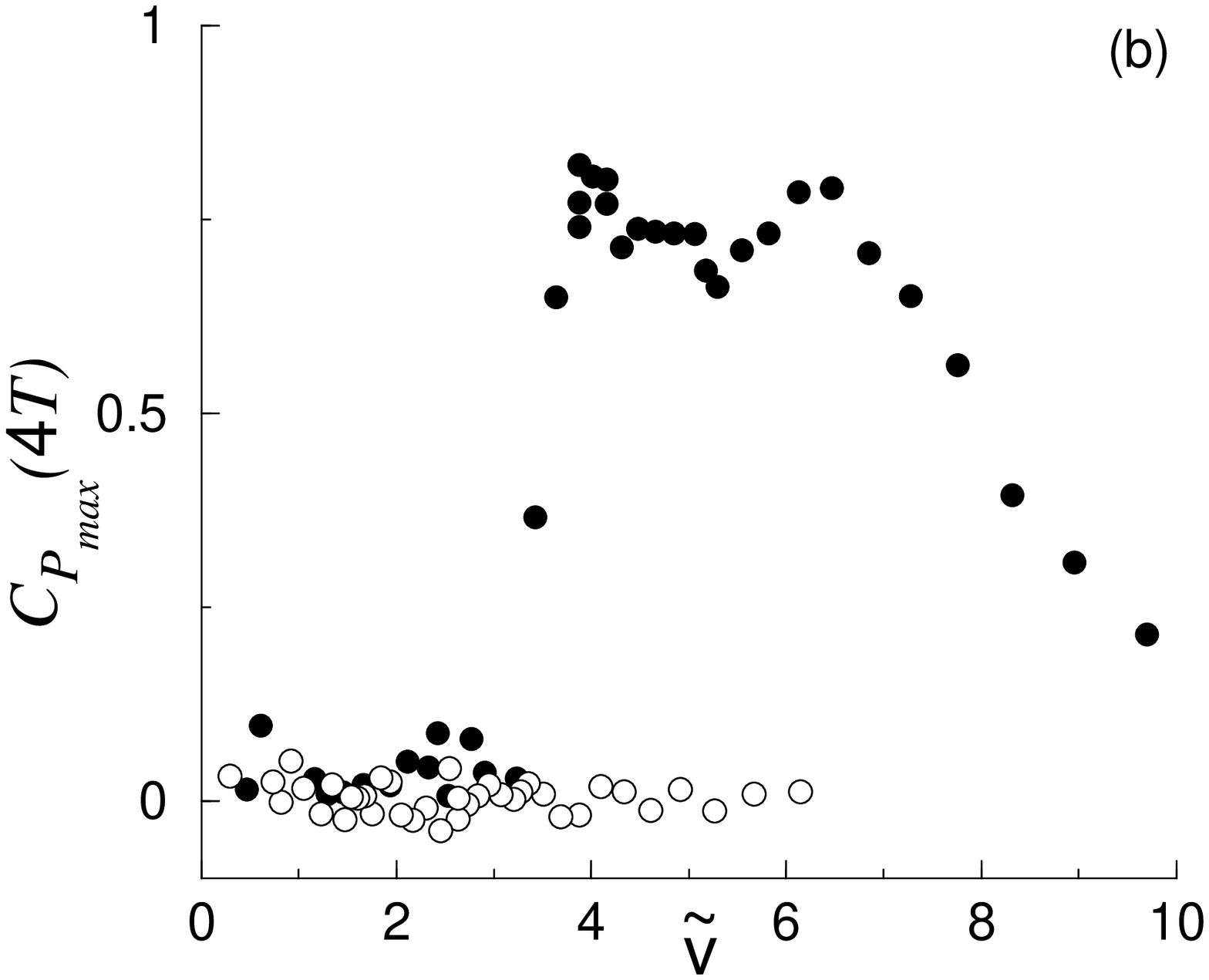}
  \caption{Grain mobility below wave onset at $\Gamma=1.9$ ({\large
  $\circ$}) and in wave regime at $\Gamma=3.0$ ($\bullet$),
  characterized by (a) pressure fluctuations and (b) autocorrelation
  of peak pressure at a delay of 4T ($N=9$).  Near
  $\tilde{\mathrm{v}}=3$, both quantities change, which
  marks the grain mobility transition.}  \label{fig13}
\end{figure}

To quantitatively investigate the proposed grain mobility transition,
fluctuations in the local collision pressure are studied.  In the
presence of waves and for $\tilde{\mathrm{v}} >
\tilde{\mathrm{v}}_{gm}$, relatively large fluctuations in the
collision pressure associated with changes in the local layer height
due to grain motion are expected.  Conversely, for $\tilde{\mathrm{v}}
< \tilde{\mathrm{v}}_{gm}$ fluctuations should be smaller because the
layer height is essentially constant.  Figure \ref{fig13}(a) plots the
relative standard deviation of the maximum pressure $\sigma_{P_{max}}
= \sqrt{(\Delta P_{max})^2}/ \langle P_{max} \rangle$ versus
$\tilde{\mathrm{v}}$ for both the flat-layer and wave regimes where
$\sqrt{(\Delta P_{max})^2}$ and $\langle P_{max} \rangle$ are the
standard deviation and mean of $P_{max}$ respectively.  For the
flat-layer, $\sigma_{P_{max}}$ is small and nearly independent of
$\tilde{\mathrm{v}}$.  For the wave regime, $\sigma_{P_{max}}$ shows
three distinct behaviors --- $\tilde{\mathrm{v}} < 3$:
$\sigma_{P_{max}}$ is independent of $\tilde{\mathrm{v}}$ and equal to
the flat-layer value; $3 < \tilde{\mathrm{v}} < 7$: $\sigma_{P_{max}}$
increases with increasing $\tilde{\mathrm{v}}$ and is larger than the
flat-layer value; $\tilde{\mathrm{v}}> 7$: $\sigma_{P_{max}}$
decreases with increasing $\tilde{\mathrm{v}}$.

As a further check that the equality of $\sigma_{P_{max}}$ for the
flat-layer and waves for $\tilde{\mathrm{v}} < 3$ is due to a loss of
horizontal grain motion, Fig.~\ref{fig13}(b) presents measurements of
the autocorrelation of $P_{max}$ at a delay of 4T, $C_{P_{max}}(4T)$
\cite{acdiss}. $C_{P_{max}}(4T)$ is sensitive to periodic variations
in $P_{max}$ even when the intrinsic pressure noise is larger than the
fluctuations associated with grain motion.  As is true for
$\sigma_{P_{max}}$, $C_{P_{max}}(4T)$ for the flat-layer is
independent of $\tilde{\mathrm{v}}$.  For waves, the dependence of
$C_{P_{max}}(4T)$ on $\tilde{\mathrm{v}}$ can also be divided into
three regimes --- $\tilde{\mathrm{v}} < 3$ : $C_{P_{max}}(4T)$ is
constant and equal to the flat-layer value; $3 < \tilde{\mathrm{v}} <
7$ : $C_{P_{max}}(4T)$ is constant and larger than the flat-layer
value; $\tilde{\mathrm{v}} > 7$ : $C_{P_{max}}(4T)$ decreases with
increasing $\tilde{\mathrm{v}}$.  The transition in $C_{P_{max}}(4T)$
to the flat-layer value is sharp and occurs at the same
$\tilde{\mathrm{v}}$ where $\sigma_{P_{max}}$ for waves reaches its
minimum value.  Also, note that $\sigma_{P_{max}}$ peaks at the
velocity ($\tilde{\mathrm{v}}=7$) at which $C_{P_{max}}(4T)$ begins to
decrease for increasing $\tilde{\mathrm{v}}$.

Our interpretation of the three regimes for $\sigma_{P_{max}}$ and
$C_{P_{max}}$ shown in Fig.~\ref{fig13} is as follows.  For
$\tilde{\mathrm{v}} < 3$ horizontal grain mobility is strongly reduced
and the layer depth is everywhere equal.  Waves in this regime are
likely bending waves \cite{mmpp} since mass transfer, dilational waves
\cite{surf}, or any other mode giving rise to a periodic variation in
pressure would produce positive correlations in the pressure
fluctuations.  For $\tilde{\mathrm{v}} > 3$ waves are mass transfer
waves. We speculate that the decrease in $\sigma_{P_{max}}$ and
$C_{P_{max}}$ above $\tilde{\mathrm{v}}=7$ results from the disordered
patterns found in this regime which possibly result from an
instability associated with the more rapid growth of the wave
amplitude in comparison to $\lambda$ as $f$ decreases \cite{pconst}.

\begin{figure}[t]  
  \centering \includegraphics[width=0.57\textwidth]{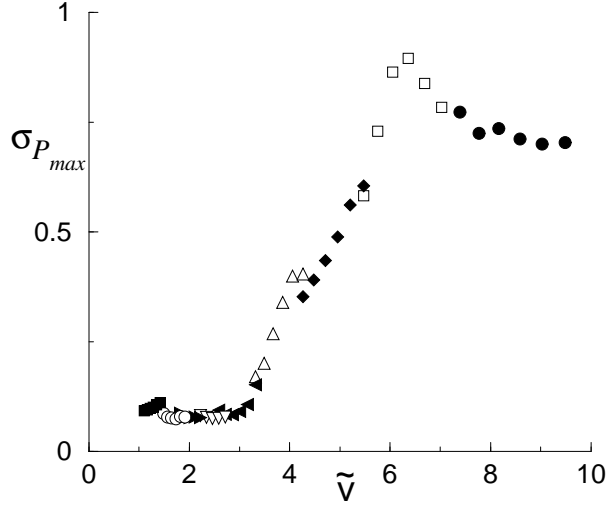}\\
  \caption{Normalized collision pressure fluctuation in wave regime
  versus dimensionless peak container velocity for varying $\Gamma$
  and $f \sqrt{D/g} = 0.05$ ($\bullet$), 0.07 ($\square$) , 0.09
  (filled {\large $\diamond$}), 0.11 ($\triangle$), 0.14 (filled
  {\large $\triangleleft$}), 0.17 ({\footnotesize
  $\bigtriangledown$}), 0.21 (filled {\large $\triangleright$}), 0.25
  ({\large $\circ$}), and 0.33 (filled $\square$) ($N=9$).  Note the
  similarity with Fig.\ref{fig13}(a).}  \label{fig14}
\end{figure}

Having made the case for a grain mobility transition in terms of
$\sigma_{P_{max}}$ and $C_{P_{max}}$, we now examine the dependence of
the transition on $\mathrm{v}$, $N$, and $D$.  Figure \ref{fig14}
presents $\sigma_{P_{max}}$ for varying $\Gamma$ and nine distinct $f$
values ranging from $0.05\sqrt{g/D}$ to $0.33\sqrt{g/D}$, as a
function of $\tilde{\mathrm{v}}$. Most importantly, Fig.~\ref{fig14}
shows that variations in $\Gamma$ at constant $f$ produce the same
changes in $\sigma_{P_{max}}$ as do variations in $f$ at constant
$\Gamma$ (see Fig.~\ref{fig13}(b)).  This finding strengthens our
proposition that $\tilde{\mathrm{v}} = \Gamma/(2 \pi f \sqrt{D/g})$ is
the correct dimensionless parameter specifying the grain mobility.
The data also show that $\tilde{\mathrm{v}}_{gm} \approx 3$, which is
in agreement with the value obtained from the pressure fluctuation
data for $\Gamma=3.0$.

\begin{figure}[tb]  	
  \centering \includegraphics[width=0.5\textwidth]{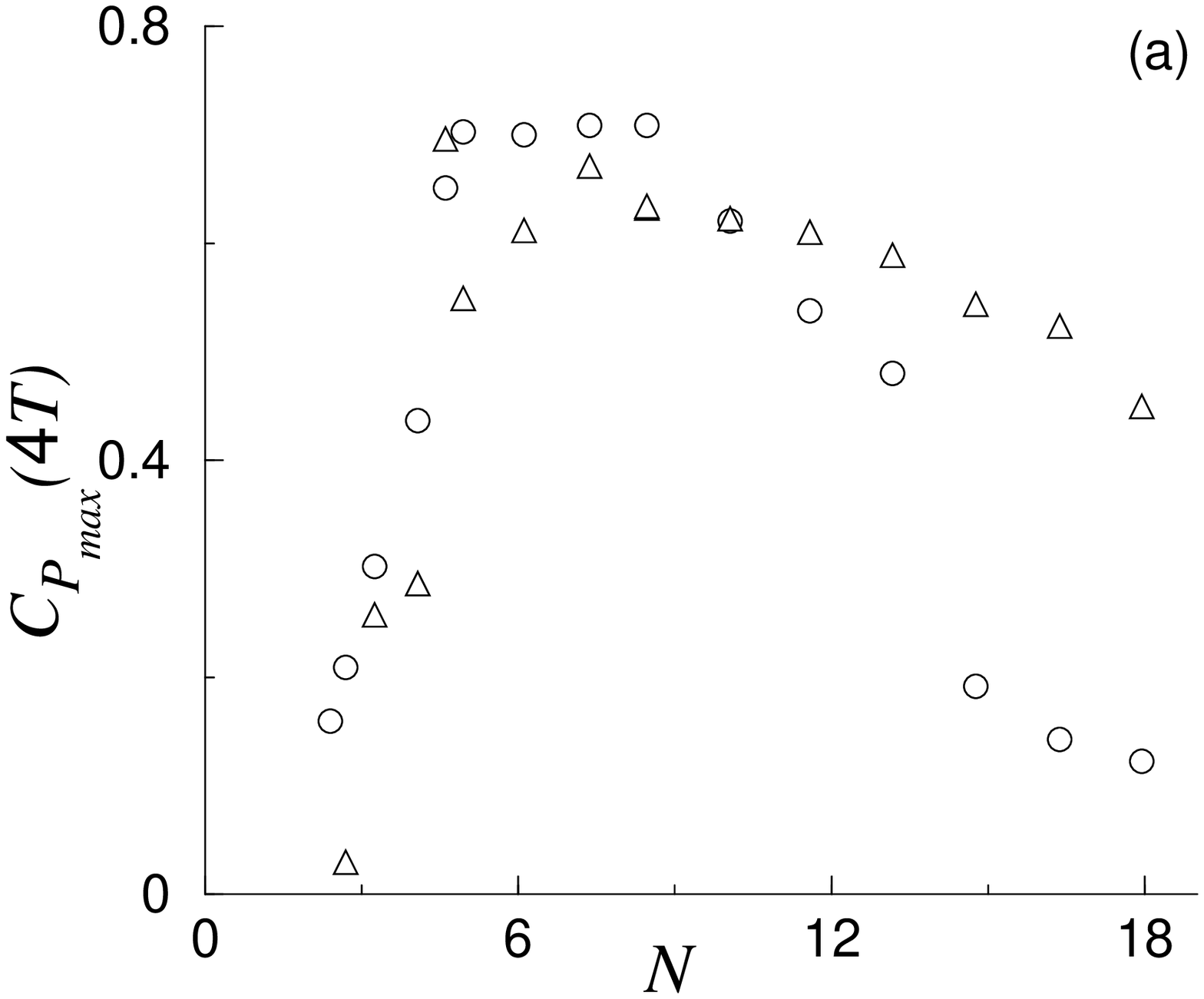}%
  \hfill \includegraphics[width=0.5\textwidth]{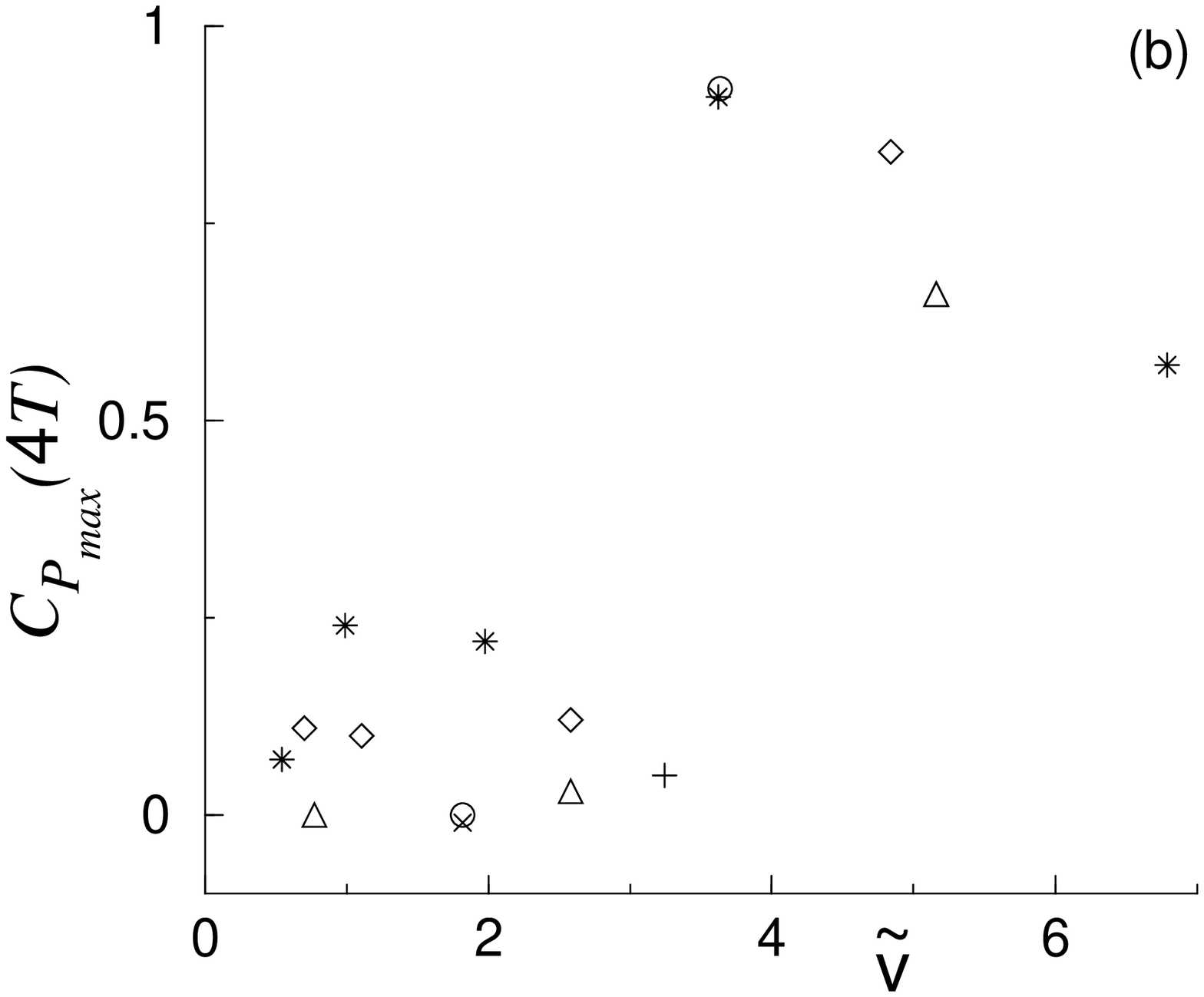}
  \caption{(a) Autocorrelation of peak pressure versus $N$ showing
  grain mobility transition at $N=14$ for $\tilde{\mathrm{v}} = 3.4$
  ({\large $\circ$}) and absence of transition for
  $\tilde{\mathrm{v}}=4.2$ ($\triangle$) ($f=30$ Hz).  (b)
  Autocorrelation of peak pressure versus $\tilde{\mathrm{v}}$
  illustrating the independence of grain mobility on $D$ for $D$ and
  $N$ of $0.66$ mm, $9$ ({\large $\circ$}); $0.66$ mm, $4$ ($\times$);
  $0.33$ mm, $9$ ({\large $\diamond$}); $0.33$ mm, $4$ ($\triangle$);
  $0.46$ mm lead, $3$ ($+$); and $0.17$ mm steel, $9$ ($\ast$).}
  \label{fig15}
\end{figure}

Figure \ref{fig15}(a) examines the dependence of grain mobility on
layer depth by plotting $C_{P_{max}}(4T)$ versus $N$ for
$\tilde{\mathrm{v}}=3.4$ and $\tilde{\mathrm{v}}=4.2$.  By choosing
$\tilde{\mathrm{v}}$ slightly above $\tilde{\mathrm{v}}_{gm} \approx
3$ (for $N=9$), $C_{P_{max}}(4T)$ is sensitive to the grain mobility
transition.  Below $N=14$, the layer response for the different
$\tilde{\mathrm{v}}$ is similar.  For $N<5$, $C_{P_{max}}(4T)$
decreases to near $0$ at $N \approx 2$, while for $5 < N < 14$,
$C_{P_{max}}(4T)$ decreases slightly with increasing $N$.  The
dependence of $C_{P_{max}}(4T)$ on $N$ at small $N$ appears similar to
that of $C_{P_{max}}(4T)$ on $\tilde{\mathrm{v}}$ at large
$\tilde{\mathrm{v}}$, suggesting that the two limits of large
$\tilde{\mathrm{v}}$ and small $N$ are related.  For $N > 14$,
$C_{P_{max}}(4T)$ continues to decreases slowly with increasing $N$
for $\tilde{\mathrm{v}}=4.2$, but, for $\tilde{\mathrm{v}}=3.4$ and $N
\geq 14$, $C_{P_{max}}(4T)$ drops to approximately the flat layer
value.  Possibly, the reduction in grain mobility with increasing $N$
is due to a decrease in grain velocity associated with an increase in
the grain collision frequency.  A more mundane explanation is as $N$
increases, the layer mass becomes significant with respect to the
container mass.  This reduces the change in layer velocity at impact
and subsequently decreases the effective driving.

Finally, Fig.~\ref{fig15}(b) looks at the effect of varying $D$ on the
grain mobility by plotting $C_{P_{max}}(4T)$ versus
$\tilde{\mathrm{v}}$ for $D =0.17$, 0.33, 0.46, and 0.66 (pressure
data for constant $\tilde{\mathrm{v}}$ and varying $N$ was not
collected).  The dependence of $C_{P_{max}}(4T)$ on
$\tilde{\mathrm{v}}$ and the value of $\tilde{\mathrm{v}}_{gm} \approx
3$ is in accord with the data for constant $D$ and $N$ and varying $f$
shown in Fig.~\ref{fig13} and is consistent with our proposal that
$\tilde{\mathrm{v}}_{gm}$ is independent of $D$. In addition to
bronze, Fig.~\ref{fig15}(b) also includes data for layers of lead and
steel particles.

In summary, $\tilde{\mathrm{v}}_{gm}$ values obtained from local
pressure measurements and from identifying the breakdown in wavelength
scaling with $H$ are in good qualitative agreement.  Both show the
same weak dependence on $N$ for small $\Gamma$ and are independent of
$D$.  Quantitatively, the value of $\tilde{\mathrm{v}}_{gm}$ for $N <
12$ obtained from the dispersion data ($\approx 2.6$) is somewhat
smaller that the value obtained from the local pressure fluctuations
($\approx 3$).  A possible explanation for this discrepancy is the
layer ``freezes'' from the bottom up as $\tilde{\mathrm{v}}$ is
reduced.  For $ 2.6 < \tilde{\mathrm{v}} < 3.0$ the upper portion of
the layer may be mobile but the associated mass variations are not
seen in the pressure fluctuations because the variation in $P$ is
small relative to the inherent pressure noise associated with a flat
layer and/or stress chains in the frozen portion of the layer wash out
the pressure fluctuations by distributing them over an area comparable
to or larger than $\lambda$.


\section{Conclusion}

Two length scales, the layer depth $H$ and the particle diameter $D$,
control the scaling properties of granular waves and granular wave
patterns.  As a guide, Table 2 provides a summary of the relevant
dimensionless parameters and their associated transition values and
figures.  The horizontal mobility of grains appears to be determined
by the container velocity --- above a critical value $\mathrm{v}_{gm}
\approx 3 \sqrt{Dg}$, grains are mobile while below $\mathrm{v}_{gm}$,
the layer depth remains uniform throughout the oscillation cycle.  The
physical mechanism responsible for the transition remains to be
understood.  The sharpness of the grain mobility transition stands in
contrast to the gradual change in properties of gases as the length
scale of disturbances approaches the mean free path.  Additionally, it
is surprising that finite grain size effects are made manifest via
$\tilde{\mathrm{v}}$ rather than $H/D$ or $\lambda/D$ as mentioned in
Sec.~\ref{wlscaling}.

\begin{table}
  \centering
  \caption{Summary of parameters, transitions and associated figures.}
  \begin{tabular}{l l c c} Symbol&Description&Value&Figures\\ \hline
  \hspace{0.1in} $\Gamma$ & Dimensionless Acceleration & $4 \pi^2 A f^2/g$ &--- \\
  \hspace{0.1in} $\Gamma_{c}$ & Flat-layer/Waves Transition & $2.2$ ($\Gamma\downarrow$) & 3--5 \\
  \hspace{0.1in} $f^*$ & Dimensionless Frequency & $f\sqrt{H/g}$ & 5, 8, 9 \\
  \hspace{0.1in} $f^*_{ss}$ & Square/Stripe Transition&0.33 & 1, 6, 7\\
  \hspace{0.1in} $\tilde{\mathrm{v}}$ & Dimensionless Velocity & $2 \pi A f /
  \sqrt{Dg}$ & --- \\ \hspace{0.1in} $\tilde{\mathrm{v}}_{gm}$ & Grain Mobility
  Transition &$3$ ($N<13$)&10--15\\
\end{tabular}
\end{table}

For $\mathrm{v}>\mathrm{v}_{gm}$, the layer behaves as a continuum
with $\lambda$ and $f$ scaling with $H$ and $\sqrt{H/g}$,
respectively.  With this scaling, dispersion curves for different
layer depth and particle size collapse onto the curve given by
Eq.~\ref{dispeqn}.  The square/stripe transition for most parameter
values occurs within the continuum regime ($\mathrm{v} >
\mathrm{v}_{gm}$), depends simply on the layer depth $H$
($f^*_{ss}=f_{ss}\sqrt{H/g} \approx 0.33$), and shows no systematic
dependence on $\Gamma$, $N$, $D$, or particle composition (for lead,
bronze, and steel).  We do not yet understand why $f^*_{ss} = 0.33$,
but speculate it is perhaps related to the dynamic angle of repose.
For $\mathrm{v} < \mathrm{v}_{gm}$, scaling with $H$ fails. Within
this regime, local pressure measurements support the idea that waves
result from a bending of the grain layer (see Mujica {\it et al.}
Ref.~\cite{mmpp}), although there is a large discrepancy between our
value of $\tilde{\mathrm{v}}_{gm} \approx 3$ and that found by Mujica
{\it et al.} --- $\tilde{\mathrm{v}}_{gm} \approx 0.6$.  Possibly, our
measurements indicate where horizontal grain mobility stops whereas
the value reported by Mujica {\it et al.}  represents the cessation of
relative vertical motion as well.  Another interesting question is the
nature of stripe patterns above the grain mobility transition.  Does
wave scaling with $H$ exist in this regime?  Casual observations show
that stripes above the transition have shorter spatial and temporal
correlations than do stripes in the sloshing regime.

Finally, the flat-layer/waves transition occurs when the layer remains
dilated after making contact with the plate.  For decreasing $\Gamma$,
this transition occurs at a nearly constant value of $\Gamma=2.2$,
which is independent of $f$ and $N$.  Associated with the onset of
waves is a sudden decrease in the collision pressure.  Possibly the
least understood aspect of the flat-layer/waves transition is the
causative connection between the dilation and the pattern.

There remain numerous unanswered questions concerning wave patterns in
vertically oscillated granular layers.  Increased computational power,
better simulation techniques, improved measurement capabilities, and
application of Navier-Stokes-like continuum equations
\cite{shattuckpp} promise significant new insights into this system in
particular and into the dynamics of granular media in general.

\section*{Acknowledgments}

The authors thank Chris Bizon and Mark Shattuck for helpful comments,
Dan Goldman for experimental assistance, and Francisco Melo for
illuminating discussions.  This research was supported by the
Engineering Research Program of the Office of Basic Energy Sciences of
the U.S. Department of Energy and by the U.S. National Science
Foundation Division of International Programs (Chile).

\vfill
\eject

\begin{thebibliography}{1}
\bibitem{smallsens} Although the sensor diameter is less than
$\lambda$ for most of the data reported in Sec.~\ref{lpress}, repeated
runs over different and/or time varying patterns allow effective
spatial averages to be obtained which are in quantitative agreement
with global pressure measurements from Refs.~\cite{surf,mmpp}.
\bibitem{pnote} In all experimental runs where $\Gamma$ is varied,
$d\Gamma / dt \approx 0.002 f$, a rate at which the measured
quantities are unchanged by further reduction of $d\Gamma/dt$.
\bibitem{crtgam} One might also argue that above $f^* = 0.4$
$\Gamma_{c\downarrow}$ and $\Gamma_{c\uparrow}$ exhibit the same
dependence on $f$ but with an offset.  If correct, this observation
may be related to the solid-like properties of the wave state above
the grain mobility transition (see Sec.~\ref{gmt}).
\bibitem{minN} For $\Gamma=3.0$, no obvious departure from the
collapse curve is evident for $N>11$ since $\lambda$ is near its
limiting value of $H$.
\bibitem{pconst} The constant value of $P^*_{max}$ for
$\tilde{\mathrm{v}} > 2.6$ can be interpreted as implying that the
pattern amplitude $A_p$ is proportional to $1/f^2$.  Writing
$P^*_{max} \propto v_{coll}/g t_{coll}$ and setting $t_{coll} = A_p /
v_{coll}$ gives $A_p \propto v^2_{coll} / g P^*_{max} \propto 1/f^2$
since $P^*_{max}$ is constant.
\bibitem{thsref} A related idea for grain mobility was proposed in
Ref.~\cite{umbths} in terms of the average layer dilation per
particle.
\end{thebibliography}
\end{document}